# Document Summarization

# with applications to Keyword extraction and Image Retrieval

**M.Tech Dissertation**

*Submitted in partial fulfillment of the*

*requirements for the degree of*

**Master of Technology**

by

**Jayaprakash S**

under the guidance of

**Prof. Dr. Pushpak Battacharya**

Department of Computer Science and Engineering

Indian Institute of Technology, Bombay

Mumbai

# Acknowledgement

I would like to thank my guide, Prof. **Dr. Pushpak Battacharya** for the consistent support and guidance he provided throughout the semester.

I would thank **Mr. Krishanu Seal**, **Mrs. Prateeksha Chandraghatgi**, **Mrs. Kiran Pulukarni** and **Mr. Muthusamy Chelliah** from Yahoo!, India for giving opportunity work with them and their constant support. Finally I would like to thank entire CLIA team, Prof. Ganesh Ramakrishana, Mr. Arjun Atreya, Mr. Swapnil, and Mr. Biplab Das for valuable guidance and discussions.

**Abstract**

Automatic summarization is the process of reducing a text document in order to generate a summary that retains the most important points of the original document. In this work, we study two problems - i) summarizing a text document as set of keywords/caption, for image recommedation, ii) generating opinion summary which good mix of relevancy and sentiment with the text document.

Intially, we present our work on an recommending images for enhancing a substantial amount of existing plain text news articles. We use probabilistic models and word similarity heuristics to generate captions and extract Key-phrases which are re-ranked using a rank aggregation framework with relevance feedback mechanism. We show that such rank aggregation and relevant feedback which are typically used in Tagging Documents, Text Information Retrieval also helps in improving image retrieval. These queries are fed to the Yahoo Search Engine to obtain relevant images [1]. Our proposed method is observed to perform better than all existing baselines.

Additonally, We propose a set of submodular functions for opinion summarization. Opinion summarization has built in it the tasks of summarization and sentiment detection. However, it is not easy to detect sentiment and simultaneously extract summary. The two tasks conflict in the sense that the demand of compression may drop sentiment bearing sentences, and the demand of sentiment detection may bring in redundant sentences. However, using submodularity we show how to strike a balance between the two requirements. Our functions generate summaries such that there is good correlation between document sentiment and summary sentiment along with good ROUGE score. We also compare the performances of the proposed submodular functions.

[1]http://images.search.yahoo.com/images

# List of Figures

# List of Tables

# List of Algorithms

# List of Abbreviations

| **Acronym** | **W**hat (it) **S**tands **F**or |
|---|---|
| **ML** | **M**achine **L**earning |
| **NLP** | **N**atural **L**anguage **P**rocessing |
| **IR** | **I**nformation **R**etrieval |
| **NB** | **N**aive **B**ayes |
| **SVM** | **S**upport **V**ector **M**achines |
| **SA** | **S**entiment **A**nalysis |

# Contents

# Chapter 1

# Introduction

...a wealth of information creates a poverty of attention...

*Herbert A. Simon*

## 1.1 Problem Statement

In this project, we are trying to extract keywords from a given news article. Intent of doing keywords extraction is to retrieve an relavant image for a given news article without annotated image. So the extracted keywords should closely match with descriptive metadata of an relevant image. After extracting keywords from text, keywords will be used for retrieving an image using Image Search System or Engine.

## 1.2 Motivation

We live in Information age in which information is freely exchanged and knowledge is easily accessed. Information is represented in several formats like text, images, videos, etc. More importantly, each of these representation convey information to the users at different rate and an image likely to provide instance sense of contentment than text.People are genetically wired to respond differently to visuals than text. For example, humans have an innate fondness for images of wide, open landscapes, which evoke an instant sense contentment.

The motivation behind finding an relevant image for a news article are

- **Increased user satisfaction**

  If a image is attached with news text, by looking at the image user can guess about the story in that news item. Hence avoids reading text to make decision on whether to read it or not. Further a better image may induce user to read the news article.

- **Content colloboration**

  Linking different representation of content coveys same infromation.

## 1.3 Contributions of this work

### Deliverables

- Image Recommendation System based on Rank Aggregation Framework is hosted in Yahoo internal site http://mediumtedium.corp.sg3.yahoo.com:4080/final.jsp.
- Rank Aggregation Framework - Image recommendation system. This is hosted in CFILT[1] server http://10.144.22.120:8040/os/.

Contibutions towards this project includes:

As keyphrase extraction is very important for this problem, we studied the existing works available on unsupervised (TextRank, RAKE) and supervised (KEA) approaches. Our contribution related to keyword extraction includes,

- Caption Generation - Generating Cpation for a given articles using probabilistic models.
- Rank Aggregation Framework - Image recommendation system which get final keyword list combined from different systems with relevance feedback as image description.
- Unsupervised Approach based on counting and boosting after co-reference. Modified Text-Rank algorithm utilizing the co-reference resolution. Supervised approach based on naive bayes classification achieves 20% overlap with the given metadata.

With respect to opinion summarization, our contribution includes the formulation of submodular functons, implementation of systems, and evaluation.

[1] Center for Indian Language Technology, IIT Bombday

## 1.4 Challenges

Challenges of finding a relevant image for a news article are,

- News document is large in size whereas metadata has very few words. Searching by entire document text may not be viable and it may be not fetch an image.
- Scalability of content-based image retrieval systems.
- It is difficult to decide the number of keywords should be used for image retrieval.Less number of keywords may fetch irrelevant images and more keywords may not fetch any images.
- Even to decide on whether suggested image is relevant to article or not.

Challenges of summarizing a document are,

- Making sure that generated summary conveys as much as information from document. In short theme of document should not be missed.
- **Diversity and Aspect Coverage**
  A produced summary should be diverse enough, and should cover information about almost all aspects.
- **Sentiment Preservation**
  Making sure that sentiment of summary matches with document summary.

## 1.5 Organization of Dissertation

The remaining part of this report organized as follows:

- Chapter 2 covers literature study and background on keyword extraction, image retrieval and opinion summarizaion.
- In Chapter 3, we describe our system for keyword extraction and image retrieval. Few unsupervised keyword extraction methods and probablisitc models for caption generation are discussed. Further Rank Aggregation Framework also discussed where keywords from different systems are combined.

- Chapter 4 describes the experimental setup, results, and example for formulation and system explained in Chapter 3.
- Chapter 5 introduces submodularity and monotone submodular functions for subset selection problem. [17] work on using submodular function for extractive summarization on multi-document is discussed.
- In Chapter 6, we present our work on opinion summarization using sub modular functions. We discuss five formulations which are montone submodular which can used aspect coverage and relevance scoring of summary sentences. Results, Case studies, and System are also discussed in the chapter.
- Chapter 7 concludes our work and also gives future implications coming out of this work.

# Chapter 2

# Literature Survey and Background

In this chapter, we present the literature survey done for our work with Opinion Summarization, Keyword Extraction and Image Recommendation. The road map of this chapter is as follows. In Section-2.1, we introduce opinion summarization and describe existing opinion summarization techniques. In Section-2.2, we describe existing work on keyword extraction techniques and image retrieval methods.

## 2.1 Keyword Extraction and Image retrieval Techniques

Unsupervised keyword extraction methods mostly rely on the relationship between the words in the text. Importance of the word is estimated based on exploiting the relationship to all other words.

### 2.1.1 TextRank

TextRank is a graph based unsupervised ranking algorithm formulated for text processing via extracting keywords and sentence extraction from documents. This method is based on Graph-based ranking algorithms like HITS algorithm [29] , Google Search Engine's PageRank [29] where is used for social networks, citation analysis and analysis of link structure world wide web. These graph based algorithms exploit the global information computed iteratively rather than looking at only surrounding information.

The basic idea of the graph based ranking models is 'voting' or 'recommendation'. If one vertice connected to another vertex, basically it is casting a vote for that other vertex. The higher number of votes the one vertex gets, the same amount of importance it gets. The importance of the vertex which is casting vote is going to determine the amount of importance should be given to that vote. Effectively the score associated with the vertex is determined by votes that are cast for it and scores of vertices that are casting vote for it.

Formally, considering G = (V, E) as directed graph with set of vertices V and set of edges E, where E is the subset of VxV. Let $In(V_i)$ be the set of vertices that are voting $V_i$ (predecessors) and $Out(V_i)$ be the set of vertices that $V_i$ is voting to. The score of vertex $V_i$ is calculated as follows [29],

$$S(V_i) = (1 - d) + d \sum_{j \in In V_i} \frac{1}{|Out(V_j)|} * S(V_j) \tag{2.1}$$

here d is called as damping factor that can be set between 0 and 1 initially, which has the role of adding a probability directly jumping from one vertex to another vertex (actually it signifies the default or implicit voting given by any vertex to all other vertices to avoid dead-end in random walk).

Starting from arbitrarily assigned values to all vertices in the graph, the vertex values are computed iterated until convergence or given threshold is reached. After convergence is reached, the scores of the vertices represents the importance of the vertices within the graph.

**Document as graph**

To use the graph ranking algorithm for natural language text, first we should build a graph that represent the given text and interconnected words and relations between them. Based on the application, text units of various sizes can be used as vertices (examples words, sentences, phrases etc). We can decide type of relations should exist between vertices e.g. lexical or semantic relations, contextual overlap *etc* based application.

Overall steps of this algorithms is,

1. Identify the text units of document and add them vertices to the graph.

2. Figure out the relations between text units, that best suits the application. Edges that connects the vertices can be undirected or directed, Weighted or Unweighted.

3. Assign the initial scores of vertices arbitrarily and iterate through ranking algorithm until convergence.

4. Sort the vertices based on final score. and selects top-K vertices as candidate vertices or text units.

5. [Optional ]Post processing is applied to vertices or textual units.

The expected result of keyword extraction task is set of keywords or phrases for a given natural language text. Any relation between two lexical units can be used as connection between two vertices. Here in this paper, co-occurrence relation with the controlled distance is used as edges or connection between vertices. Two vertices are said to be connected if that two lexical units tend to co-occur within a window of N words, where 'N' can vary.

The vertices added to graph contains lexical units of certain types, for instance in this paper [25] they have used individual words as the vertices and connection between vertices (individual words) represents that they co-occurred in the text within the window size N.

**Undirected Edges**

Graph used for ranking keyword is undirected, whereas original algorithm was developed for directed graphs. If the two words tend to co-occur then they are mutually connected to each other, so each vetices in-degree equal to out-degree.

**Weighted Edges**

Edges in the TextRank model is weighted, they directly indicate the strength of connection between two vertices. In this if we two words tend to co-occur frequently then they will have strong (more weight) connection.

By considering above undirected and Weighted cases, the original graph based ranking algorithm has been modified into as follows,

$$WS(V_i) = (1-d) + d * \sum_{j \in In V_i} \frac{W_{ji}}{\sum_{k \in Out V_j W_{jk}}} * WS(V_j) \qquad (2.2)$$

**Example-Text**

Apple's product road map is a topic that may receive more speculation than any subject in all of tech. With that in mind, some are expecting 2014 to be a very big year for the Cupertino, Calif.-based maker of the iPad and iPhone.

Jefferies analyst Peter Misek, he of the precarious Apple upgrade, says 2014 will indeed be a crucial year for Apple as the company lays out its next version of the iPhone, the iPhone 6.

Misek notes that the next phone will likely have a new design and a much bigger screen than its predecessor. A 4.8-inch screen is likely size. The iPhone 5s/5c has a 4-inch screen. 'We discovered from Asian players that Apple is aggressively investing in OLED alongside its display partners,' Misek wrote in his note. 'Apparently Apple has begun to procure equipment for LG Display, Sharp, and Japan Display.'

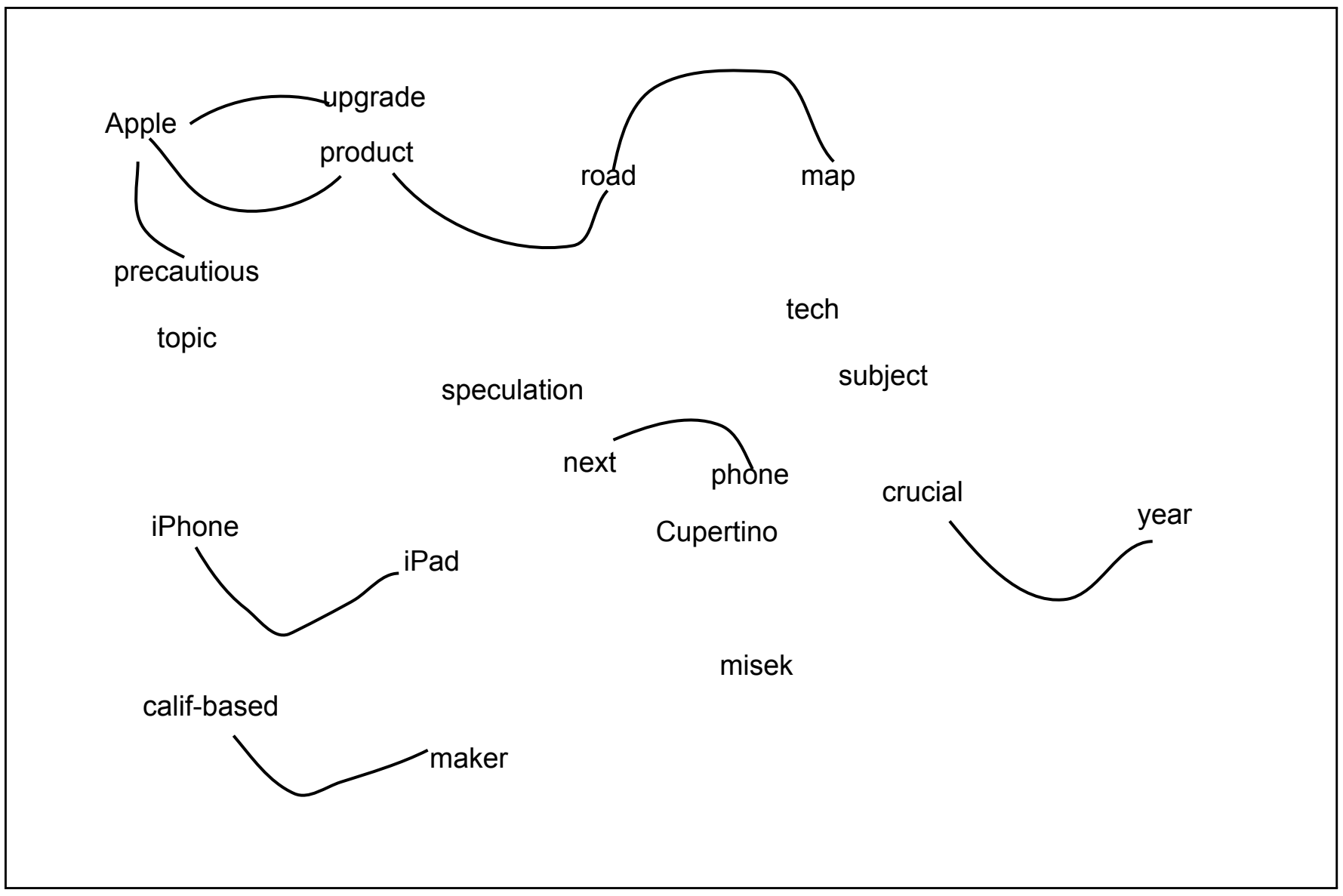

Figure 2.1: TextRank - Graph for text document

**Process**

First the given document is tokenized into words, and syntatic classes of each word (part of speech tag) is identified. It is said that picking only certain syntatic classes gives the better precision (nouns and adjectives). Only unigram words considered as vertices. Graph ranking algorithm run on the constructed graph. Top fraction of vertices selected based on score given to vertices on convergence. If the selected unigram words tend to co-occur in the text, they are combined together and considered as multi-word keywords or keyphrases.

**Results**

This algorithm was tested against 500 science articles where keywords algorithm was compared with manually annotated keywords. It is shown that this algorithm achieves highest F-score 36.2% when edges are considered as undirected with co-occurence window size (N) is 2.

### 2.1.2 RAKE

Rapid Automatic Automatic Keyword Extraction (**RAKE** ) [**?** ] is unsupervised, language independent method for extracting keywords from individual documents. RAKE is based on the assumption that keywords contains multiple words at large but very rarely contain stop words and punctuation in it.

RAKE needs stop words list, phrase delimiters and word delimiters as input parameters. Candidate keywords are chosen based on the stop words and phrase delimiters. Co-occurences of words within the candidate words used as measure for candidate keyword being a keyword.

First, the document is split into array of words based on the word delimiters. The resultant array is splited into sequence of continuous words based on the phrase delimiters and stop word occurence.

**Example - Text Document**

Criteria of compatibility of a system of linear Diophantine equations, strict inequations, and nonstrict inequations are considered. Upper bounds for components of a minimal set of solutions and algorithms of construction of minimal generating sets of solutions for all types of systems are given. These criteria and the corresponding algorithms for constructing a minimal supporting set of solutions can be used in solving all the considered types of systems and systems of mixed types.

**Example - Candidate Keywords**

Compatibility - systems - linear constraints - set - natural numbers - Criteria - compatibility - system - linear Diophantine equations - strict inequations - nonstrict inequations - Upper bounds - components - minimal set - solutions - algorithms - minimal generating sets - solutions - systems - criteria - corresponding algorithms - constructing - minimal supporting set - solving - systems - systems

**Example - Final Keyword Scores**

minimal generating sets (8.7), linear diophantine equations (8.5), minimal supporting set (7.7), minimal set (4.7), linear constraints (4.5), natural numbers (4), strict inequations (4), nonstrict inequations (4), upper bounds (4), corresponding algorithms (3.5), set (2), algorithms (1.5), compatibility (1), systems (1), criteria (1), system (1), components (1),constructing (1), solving (1)

After all candidate keywords are identified and graph of co-occurences is built. Score of a candidate keyword is calculated based on sum of it's member words scores.

Word scores are based on,

- word frequency *freq(w)*
- word degree *deg(w)*
- ratio of word frequency to degree *freq(w) / deg(w)*

deg(w) favours a word which occurs in longer candidate keywords. words that occur in many candidates are favoured by freq(w). Words that largely part of longer candidate keywords are favoured by deg(w)/freq(w).

Since candidate keywords are generated based on stop words. No candidate keyword will have any stop words in it (e.g. Times of India) . So to include them as candidate keywords, If pair of words occur twice in the document and in the same order then it they are added to candidate set of keywords.

RAKE's performance is evaluated against technical abstracts reported in Hulth (2003), and it achieved 33.7 % precision and 37.2 % recall with self generated stopwords (df > 10) which is higher than textrank's best score which is 31.2 % precision and 37.2 % recall.

### 2.1.3 KEA

KEA [38] describes about the keyphrase extraction and assignment. Keyphrase extraction and assignment are statistical learning methods requires a set of documents annotated with the manually assigned keywords.

**keyphrase assignment** is to select the phrases from the controlled phrase vocabulary that best describe a document. The training data mapped to each phrase in the vocabulary, separate classifier is learned for each phrase. A new (test) document is given to all classifiers, the phrase of classifier which gives maximum positive score is choosen. We are not further discussing this technique because it is less relevant to the current scenario, where meta learning does not fit to the controlled vocabulary learning.

**keyphrase extraction** is designed to choose the keyphrase from text document itself. It is based on lexical and information retrieval techniques to extract phrases from the document text. Training data is used to tune the parameters of each features.

#### Phrase Identification

- Candidate phrases are limited to a certain maximum length (normally 3 words).
- Candidate phrases cannot be proper names.
- Candidate phrases can not start or end with stop words.

All continous sequence of words in each sentence satisfy above three rules, are candidate phrases. Subphrases also part of candidate phrases.

#### Features

The initial version of KEA used only two features for deciding importance of phrases. They are TFxIDF and first occurence of each phrase in the document.

- **TFxIDF** TFxIDF is used as one of the feature. TF is the frequency of phrase in the test document and IDF is general usage or number documents the phrase used.

$$TFxIDF = \frac{freq(P, D)}{size(D)} x - \log_2 \frac{df(P)}{N}, where \tag{2.3}$$

– freq(P,D) number of times phrase P occurs in document D.

– size(D) is the number of words in D

– df(P) is number of documents have the phrase P in total training corpus.

– N is total number of documents in collection.

- **Positional Information** First occurence of phrase in the document is used as another feature. It is calculated by number words precede the phrase's first occurence divided by the number words in the document.

Both the features are discretized. The real valued features are divided by the range they fall into and assigned categorical values.

**Classification**

Each candidate keyphrase is classified into 'YES' or 'NO' which indicates whether the candidate phrase is important or not (keyphrase or not) based on feature values of phrases.

$$P[YES] = \frac{Y}{Y+N} P_{TFxIDF}[t|YES] * P_{DISTANCE}[d|YES] \quad (2.4)$$

**Results**

KEA algorithm was test with technical abstracts of (110 training documents). 0.909 keyphrase matches on average out of 5 keyphrases extracted. 1.712 matches out of 15 keyphrases extracted.

## 2.2 Image Retrieval

An image retrieval system is a system for browsing, searching and retrieving images from a large repository of digital images. Common methods of image retrieval utilize some method of adding metadata to images such as keywords or descriptions, so that retrieval can be performed over the annotation words.

To search for images, we/user need to provide query terms such as keyword or image file and the system is expected to return images 'similar' to the query.

Image retrieval system are broadly classified as,

- Image meta search
- Content based retrieval

### 2.2.1 Image meta search

Given query as words, and the descriptive meta data of each image is considered as text document. Image search system work as traditional information retrieval system where regardless of image semantics only description of image is only used for retrieval.

### 2.2.2 Content based retrieval

If we have documents and images annotated with them. Consider D is set of documents which contain both images and text. Images and texts are represented in term of feature vectors $R^I$ and $R^T$ respectively. These vectors represented in different vector space and there exist one-to-one mapping between them. Given text $T^q \in R^T$ we need to find an $I_q \in R_I$.

For finding an image based on text, we need to learn a mapping function

$$M : R^T -> R^I \tag{2.5}$$

Given text $T^q$ it suffices to find nearest image $M(R^T)$. Since there is not direct correspondance between $R^T$ and $R^I$. The mapping has to be learned from tranining sample. One way is to map each space into intermediate spaces $U^T$ and $U^I$ such they have correspondance.

$$M_I : R^I -> U^I \tag{2.6}$$

and

$$M_T : R^T -> U^T \tag{2.7}$$

The two isomorphic spaces $U^I$ and $U^T$ and there is invertible mapping

$$M : U^T -> U^I \tag{2.8}$$

Main problem is to find the subspaces $U^T$ and $U^I$, one way is to find two linear projections

$$P_I : R^I -> U^I \tag{2.9}$$

and

$$P_T : R^T -> U^T \tag{2.10}$$

**Correlation matching**

Canonical Correlation Analysis (CCA) is a data analysis and dimensionality reduction method similar to Principle Component Analysis (PCA). Here, PCA deals with one dimension whereas CCA is joint dimensionality reduction of two e heterogeneous representations of the same data.

## 2.3 Opinion Summarization

The focus of this section is on aggregating and representing sentiment information drawn from an individual document or from a collection of documents. For example, a user might desire an at-a-glance presentation of the main points made in a single review. Another application considered within this paradigm is the automatic determination of market sentiment, or the majority "leaning" of an entire body of investors, from the individual remarks of those investors. Major application for Opinion Summarization is to provide pre-processed data for sentiment analysis task.

Opinion summarization is now an important task so as to:

- Present the user a short summary, which conveys the essence as well as the sentiment of the review.
- Provide a short subjective extract to the sentiment analysis tool for faster execution.
- Cluster and store similar documents together.

### 2.3.1 Difference with Traditional Summarization

An opinion summary is quite different from a traditional single document or multi-document summary(of factual information) as an opinion summary is often centred on entities and aspects and sentiments about them, and also has a quantative side. [21] clearly differentiated between traditional single document or multidocument summary (of factual information) and opinion summary. Traditional single document summarization produces a short text from from long text by extracting some "important" sentences. Traditional multi-document summarization finds differences among documents and discards repeated information. Neither of them explicitly captures different topics/entities and their aspects dicussed in the document, nor do they have a quantative side. The "importance" of a sentence in traditional text summarization is often defined operationally based on the summarization algorithms and measures used in each system. Opinion summarization, on the other hand, can be conceptually defined. The summaries are thus structured. Even for output summaries that are short text documents, there are still some explicit structures in them.

### 2.3.2 Aspect-Based Opinion Summarization

Chapter 1 indicated that the opinion quintuple provides the basic information for an opinion summary. Such a summary is called an aspect-based summary (or featurebased summary) and was proposed in [8] and [23]. Much of the opinion summarization research uses related ideas. This framework is also widely applied in industry. For example, the sentiment analysis systems of Microsoft Bing and Google Product Search use this form of summary.

Aspect-based opinion summarization has two main characteristics. First, it captures the essence of opinions: opinion targets (entities and their aspects) and sentiments about them. Second, it is quantitative, which means that it gives the number or percent of people who hold positive or negative opinions about the entities and aspects. The quantitative side is crucial because of the subjective nature of opinions. The resulting opinion summary is a form of structured summary produced from the opinion quintuple, as defined in Chapter 1. Figure 2.2 shows an aspect-based summary of opinions about a digital camera [8]. The aspect GENERAL represents opinions on the camera as a whole, i.e., the entity. For each aspect (e.g., picture quality), it shows how many people have positive and negative opinions

*Digital Camera* 1:

Aspect: **GENERAL**

| | | |
|---|---|---|
| Positive: | 105 | <Individual review sentences> |
| Negative: | 12 | <Individual review sentences> |

Aspect: **Picture quality**

| | | |
|---|---|---|
| Positive: | 95 | <Individual review sentences> |
| Negative: | 10 | <Individual review sentences> |

Aspect: **Battery life**

| | | |
|---|---|---|
| Positive: | 50 | <Individual review sentences> |
| Negative: | 9 | <Individual review sentences> |

…

Figure 2.2: An example of a feature-based summary of opinions

respectively. <individual review sentences> links to the actual sentences (or full reviews or blogs). This structured summary can also be visualized [23]. Figure 2.3 (A) uses a bar chart to visualize the summary in Figure 2.2. In the figure, each bar above the X-axis shows the number of positive opinions about the aspect given at the top. The corresponding bar below the X-axis shows the number of negative opinions on the same aspect. Clicking on each bar, we can see the individual sentences and full reviews. Obviously, other visualizations are also possible. For example, the bar charts of both Microsoft Bing search and Google Product Search use the percent of positive opinions on each aspect. Comparing opinion summaries of a few entities is even more interesting [23]. Figure 2.3 (B) shows the visual opinion comparison of two cameras. We can see how consumers view each of them along different aspect dimensions including the entities themselves.

The opinion quintuples in fact allows one to provide many more forms of structured summaries. For example, if time is extracted, one can show the trend of opinions on different aspects. Even without using sentiments, one can see the buzz (frequency) of each aspect mentions, which gives the user an idea what aspects people are most concerned about. In fact, with the quintuple, a full range of database and OLAP tools can be used to slice and dice the data for all kinds of qualitative and quantitative analysis. For example, in one practical sentiment analysis application in the automobile domain, opinion quintuples of individual cars were mined first. The user then compared sentiments about small cars, medium sized cars, German cars and Japanese cars, etc. In addition, the sentiment analysis results were also used as raw data for data mining. The user ran a clusteclustering algo-rithm and found some interesting segments of the market. For example, it was found that one segment of the customers always talked about how beautiful and slick the car looked and how fun it was to drive, etc, while another segment of the customers talked a lot about back seats and trunk space, etc. Clearly, the first segment consisted of mainly young people, while the second segment consisted mainly of people with families and children. Such insights were extremely important. They enabled the user to see the opinions of different segments of customers.

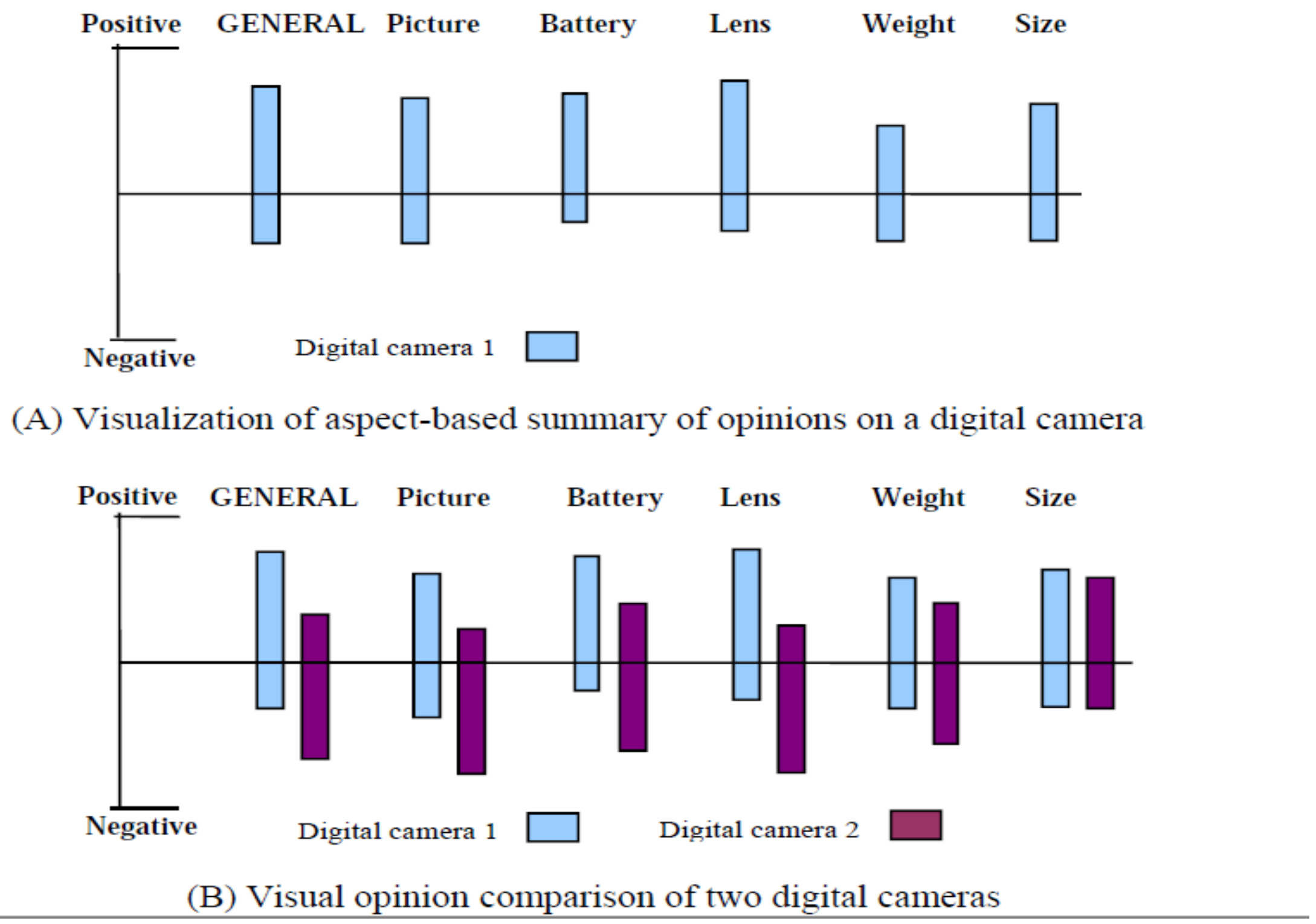

Figure 2.3: Visualization of feature-based summaries of opinions

### 2.3.3 Extractive Summarization

In [30], a mincut-based algorithm was proposed to classify each sentence as being subjective or objective. The algorithm works on a sentence graph of an opinion document, e.g., a review. The graph is first built based on local labeling consistencies (which produces an association score of two sentences) and individual sentence subjectivity score computed based on the probability produced by a traditional classification method (which produces a score for each sentence). Local labeling consistency means that sentences close to each other are more likely to have the same class label (subjective or objective). The mincut approach is able to improve individual sentence based subjectivity classification because of the local labeling consistencies. The purpose of this work was actually to remove objective sentences from reviews to improve document level sentiment classification.

[15] defined opinion summarization in a slightly different way. Given a set of documents D (e.g., reviews) that contains opinions about some entity of interest, the goal of an opinion summarization system is to generate a summary S of that entity that is representative of the average opinion and speaks to its important aspects. This paper proposed three different models to perform summarization of reviews of a product. All these models choose some set of sentences from a review. The first model is called sentiment match (SM), which extracts sentences so that the average sentiment of the summary is as close as possible to the average sentiment rating of reviews of the entity. The second model, called sentiment match + aspect coverage (SMAC), builds a summary that trades-off between maximally covering important aspects and matching the overall sentiment of the entity. The third model, called sentiment-aspect match (SAM), not only attempts to cover important aspects, but cover them with appropriate sentiment. A comprehensive evaluation of human users was conducted to compare the three types of summaries. It was found that although the SAM model was the best, it is not significantly better than others.

In [28], a more sophisticated summarization technique was proposed, which generates a traditional text summary by selecting and ordering sentences taken from multiple reviews, considering both informativeness and readability of the final summary. The informativeness was defined as the sum of frequency of each aspect-sentiment pair. Readability was defined as the natural sequence of sentences, which was measured as the sum of the connectivity of all adjacent sentences in the sequence. The problem was then solved through optimization. In [27], the authors further studied this problem using an integer linear programming formulation.

### 2.3.4 Contrastive View Summarization

Several researchers also studied the problem of summarizing opinions by finding contrastive viewpoints. For example, a reviewer may give a positive opinion about the voice quality of iPhone by saying "The voice quality of iPhone is really good," but another reviewer may say the opposite, "The voice quality of my iPhone is lousy." Such pairs can give the reader a direct comparative view of different opinions.

[11] proposed and studied this problem. Given a positive sentence set and a negative sentence set, this work performed contrastive opinion summarization by extracting a set of k contrastive sentence pairs from the sets. A pair of opinionated sentences (x, y) is called a contrastive sentence pair if sentence x and sentence y are about the same topic aspect, but have opposite sentiment orientations. The k chosen sentence pairs must also represent both the positive and negative sentence sets well. The authors formulated the summarization as an optimization problem and solved it based on several similarity functions.

## 2.4 Summary

Several researchers have also studied opinion summarization in the traditional fashion, e.g., producing a short text summary with limited or without consideration of aspects (or topics) and sentiments about them. A weakness of such traditional summaries is that they only have limited or no consideration of target entities and aspects, and sentiments about them. Thus, they may select sentences which are not related to sentiments or any aspects. Another issue is that there is no quantitative perspective, which is often important in practice because one out of ten people hating something is very different from 5 out of ten people hating something.

In this chapter, we described existing works on keyword extraction. First two unsupervised approaches TextRank, RAKE explained. Next KEA algorithm based on supervised approach is explained. In the next chapter, we describe the experiments done on keyword extraction.

# Chapter 3

# Keyword Extraction and Image Retrieval

## 3.1 Unsupervised Approaches

### 3.1.1 Boosting based on frequency and co-reference

Simple method for the determining top keywords can be based the occurrences of each word in the given document. However in natural language text a word is represented in different forms. Stemming may help normalizing different verb representation with morphemes. But images and news articles mostly centered around the entities and then verb as relations if required. *For example, Steve Jobs can be referred as Steve, Jobs, he, his, him... etc.*. When we want to find the fraction of sentences that does covers entities or nouns, we can use the co-reference resolution and anaphora resoultion. To identify the co-refered mentions in the article *Stanford Co-reference Pipeline* is used.

Steps of the experiments are,

1. Given document is tokenized in to lexical units.
2. All noun phrases from the parse tree are considered as candidate for final key phrases.
3. Stanford co-referencing pipleline ran on the sentences of articles. From the output of co-reference, the frequency of each noun phrase is calculated.
4. top noun phrases picked as candidate keywords.

**FREQUENCY BASED ON RAKING AFTER COREFERENCE**

1. **EU** CHAIN27-["the EU 's" in sentence 2, "EU" in sentence 3, "The EU 's" in sentence 4, "EU" in sentence 6, "EU" in sentence 10, "EU" in sentence 11]
2. **European Commission** CHAIN19-["The European Commission , the EU 's executive ," in sentence 2, "The European Commission" in sentence 2, "European Commission" in sentence 2, "its" in sentence 2, "its" in sentence 2, "the Commission 's" in sentence 5, "the Commission" in sentence 6, "The Commission" in sentence 7, "the Commission" in sentence 8]
3. **duties** CHAIN7-["duties on billions of euros of Chinese solar panels" in sentence 1, "the duties" in sentence 7, "the duties" in sentence 8, "the duties" in sentence 10]

**Different Weighting Methods**

Scoring methods like

- Giving more weightage to the mentions which appear in the first few sentences.
- Phrases in part of sentences,
- Referring part of phrase, considered to be talking about entity and boosted their scores. For example, in the below result, appearance of *duties* in *duties on billions of euros of Chinese solar panels* implicitly denotes we are talking about *chinese solar panels*. This needs to be studied with wide range of articles.

**RANKING USING POSITIONAL WEIGTHING**

1. **EU**
2. **European Commission**
3. **Chinese solar panels** :2:11.75: CHAIN13-["Chinese solar panels" in sentence 1, "solar panels" in sentence 11]
4. **billions** :2:9.5: CHAIN9-["billions of euros of Chinese solar panels" in sentence 1, "them" in sentence 2]
5. **duties** :4:8.0: CHAIN7-["duties on billions of euros of Chinese solar panels" in sentence 1, "the duties" in sentence "the duties" in sentence 8, "the duties" in sentence 10]

### 3.1.2 TextRank - Modified

Purpose is to tryout scoring methods other than frequency of entities or phrases. It is based on how surrounding entities are connected. We modified the TextRank [25], to take into account about co-refering mentions in the text. So, text graph is created in a different way than just considering co-occurence window.

**Creation of Graph**

- All the possible noun phrases the document used as nodes.
- Two consecutive nouns phrases in a sentences are connected. For example in the sentence 'EU will impose anti-dumping levies.', *EU* and *'anti-dumping levies'* are connected in the graph.
- Two noun phrases/entities are connected, If they are co-referred. For example consider following two sentences 'Ramu is a Student.' and 'He is intelligent.', here *Ramu* and *He* are connected.

Above three steps create undirected graph with edges are unweighted. Page Rank or Random walk applied on this graph with damping factor 0.15,

$$WS(V_i) = (1 - d) + d * \sum_{j \in InV_i} \frac{W_{ji}}{\sum_{k \in OutV_j W_{jk}}} * WS(V_j) \tag{3.1}$$

**RANKING USING MODIFIED TEXTRANK**

1. **The European Commission** , the EU 's executive, rank=1.4510955164654349, key=34, marked=true, sentNum=8, startWord=7, endWord=9, corefChain=19
2. **the EU**, rank=1.36762168208762, key=46, marked=true, sentNum=3, startWord=1, endWord=2, corefChain=28]
3. **duties on billions of euros of Chinese solar panels**, rank=1.3200079680210361, key=9, marked=true, sentNum=1, startWord=8, endWord=17, corefChain=8
4. **provisional duties**, rank=1.1529714795051218, key=42, marked=true, sentNum=11, startWord=5, endWord=7, corefChain=26]

## 3.2 Supervised Approaches

In the supervised approaches on keyword extraction, we consider the keywords are the words which appear in the meta data of image. We do remove the stop words of image meta data, all other words are considered to be keywords. and we try to learn a function which takes text document and title as input and produces the set of keywords which has maximal overlap with the keywords or metadata.

### 3.2.1 Naive Bayes

First experiment is using naive bayes classifier to find, the probability of being a word is 'keyword' or 'not'. We tokenized the text into sentences and words. Term frequency and Inverse Document Frequency is calculated after stemming the words.

*titleScore* is the probability of word being generated from title text. Often cases, title words are given importance, but difficulty is title is small in size so most of the cases the acronyms or shortened word. For example *united states* in text corresponds to *U.S.* in title, *mahindra singh doni* in text corresponds to *mahi* in title.

*IDFTF* score calculated based IR approaches.*Postag* for each word is obtained from Stanford PCFG parser.

Final score is calculated based on,

$P(Key/Word, PosTag, IDFTF, titleScore)$

$$= P(Key) * P(Word/Key) * P(PosTag/Key) * P(IDFTF/Key) * P(titleScore/Key) \tag{3.2}$$

Graphical model digram of Naive bayes approach is depicted in 3.1

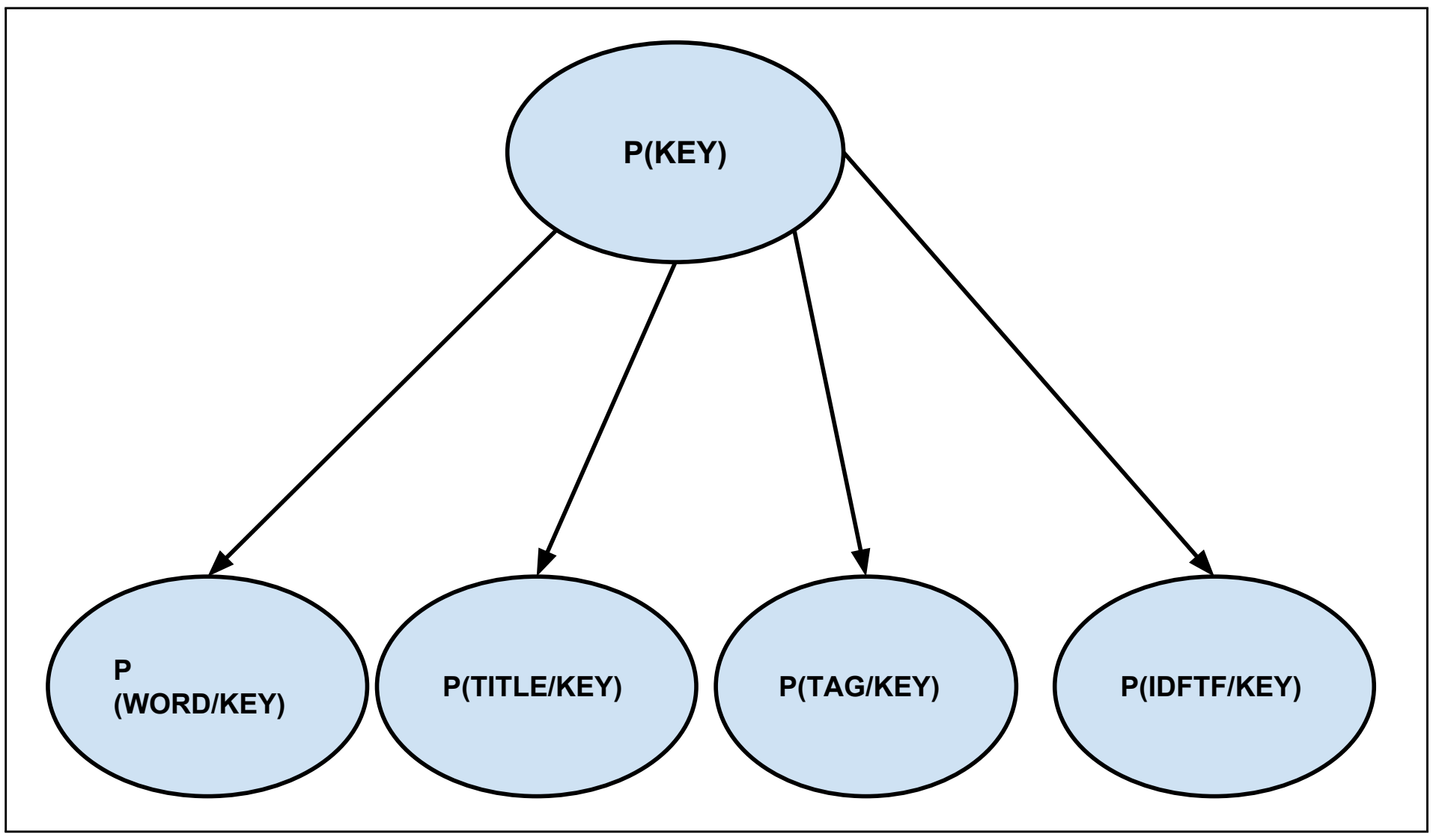

Figure 3.1: Naive Bayes Classification

We tested this method with 50 articles, its title and meta data extracted from Yahoo! News. Keywords extracted using this system achieves 20% overlap on metadata text.

In the unsupervised approaches, **Modified textrank** is performing better than **counting and proximity** based approaches. For example, in a article about *iphone's camera*, the *camera* word occurs many times in the text whereas *iphone* is not. So in the final keywords have *camera* but not iphone which is essential. Since *Modified TextRank* considers importance of sourrounding words, *iphone* also included in the keywords list.

One problem encountered with *modified textrank* algorithm is selecting long noun phrases as keyphrases. In the supervised approaches, naive bayes achieves 20% overlap with metadata.

Out finding is that none of these algorithms performs good on all the text documents. Diffirent documents need different kind of keywords and the number of keywords to be extracted also varies from document to document.

- For a document describing about a entity, one or two keywords about that entity is enough. Example: *Sachin Tendulkar, wordcup 2012*.
- For a document describing about a event, keywords should contain event, location, time *etc*. Example: *building collapse, salem, india, october 13, etc*

## 3.3 Caption Generation

All the images have caption which describes about the image, it is safe to assume that there exist one-to-one association between Caption and Image (i.e) given the proper caption, certainly we would be able to retrieve an relevant image. In this method, for a given article, We try to generate the caption using the probabilistic models.

We need to find best caption $C^*$ given a document D.

$$C^* = argmax_C P(C|D) \tag{3.3}$$

In the above equation P(C/D) is approximated into product of two parts.

$$P(C|D) \approx P(C|\{cw \in C\}) * P(\{cw \in C\}|D) \tag{3.4}$$

First part $P(\{cw \in C\}|D)$, is for choosing set of words from documents. Second part $P(C|\{cw \in C\})$, is for reordering the given set of words.

Choosing set of words cw, is based on

$$P(\{cw \in C\}|D) = \prod_{cw \in C} \sum_{dwinD} P(cw/dw) * g(dw, D) \tag{3.5}$$

where,

$P(cw/dw)$ is translation probability between caption word (cw) and document word (dw). $g(dw, D)$ is importance of word (dw), which is normalized tf-idf.

Re-ordering of chosen set words is done based on n-gram language model score.

Likelihood of caption C given D is,

$$P(C|D) \propto \frac{P(T)}{P(\{cw \in C\})} \prod_{cw \in C} \sum_{dw \in D} P(cw/dw) g(dw, D) \tag{3.6}$$

TODO:need to add the caption generated

## 3.4 Rank Aggregation Framework

We have experimented with various keyword extraction systems to produce the relavant set of terms from documents. However, Each system have its own merits works well for set of articles and not for others. We created rank aggregation system. In this rank aggregation framework, we try to choose or re-rank the terms generated from different system.

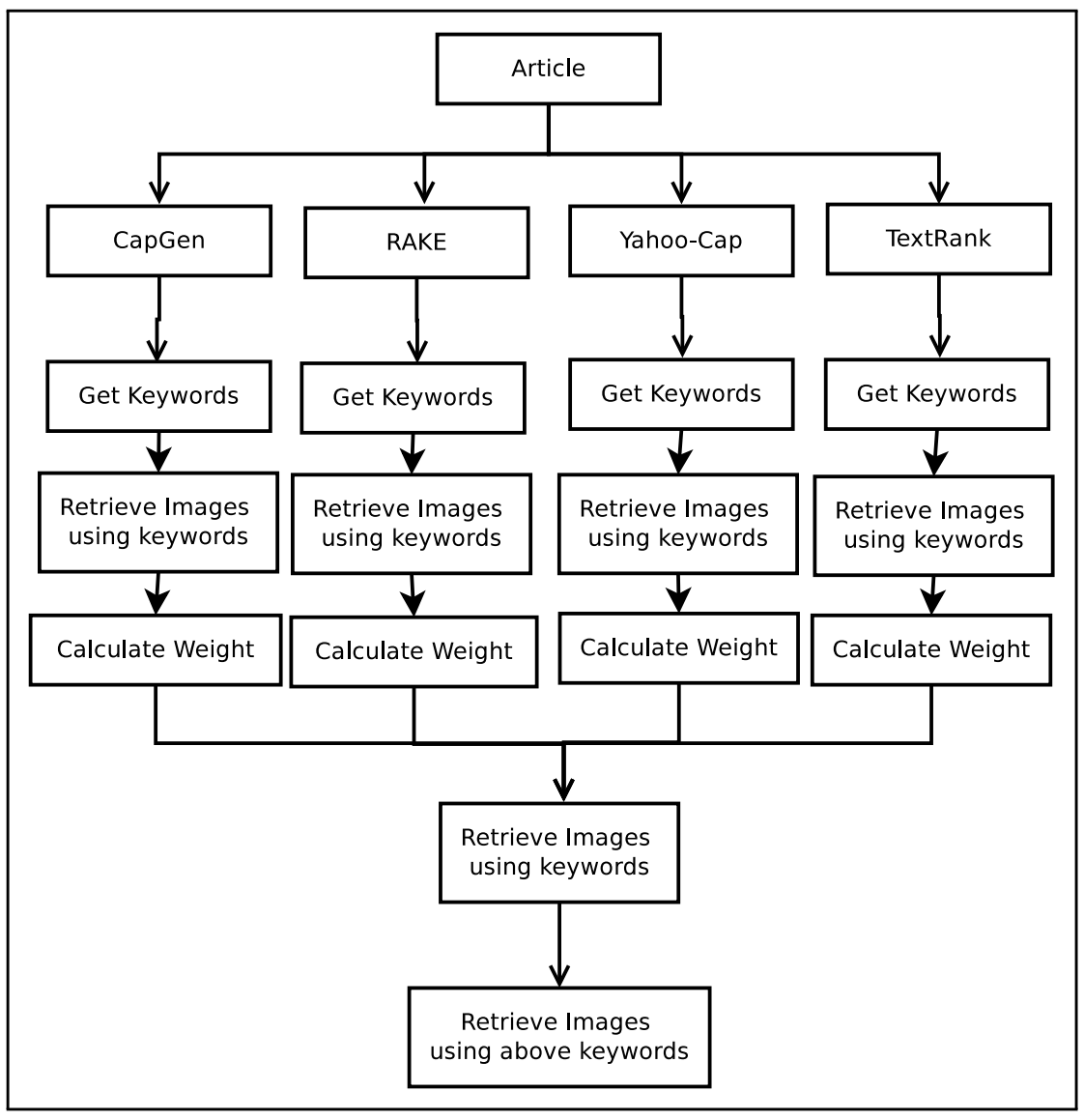

Figure 3.2: Our System - Rank Aggregation Framework for Image Recommendations

### 3.4.1 TFIDF

TFIDF (Term Frequency-Inverse document frequency), is a numeric static that basically measures the important of word to a document with respect to collection or corpus.

### 3.4.2 RAKE

RAKE begins keyword extraction on a document by parsing its text into a set of candidate keywords. First, the document text is split into an array of words by the specified word delimiters. This array is then split into sequences of contiguous words at phrase delimiters and stop word positions. Words within a sequence are assigned the same position in the text and together are considered a candidate keyword.

### 3.4.3 TextRank

TextRank [25] is Graph-based unsupervised algorithm. Graph is constructed by creating a vertex for each words in the document and edges between vertices based on the windows two words (of vertices) have in common and then, ranking them by applying PageRank [29] to the resulting graph. Summary is generated with sentences having more vertex score [25].

### 3.4.4 Yahoo-CAP

A term extraction system provided by Yahoo-Content Analysis Platform (CAP)[1].In this method, terms are ranked, which is combination of rule based and statistical based system. It has been observed that, this system works well with artciles about the entities or concepts present as part of Wikipedia in it like articles about sports, entertainment etc.

### 3.4.5 Rank Aggregation-1/KL Divergence-Uni

In this rank aggregation framework, we combine all the keywords extracted from different system to produce a better keyword set. To produce the final keyword list, we need to combine the keywords set from different systems. As we said earlier different system perform better on different articles based on article type and nature. Identifying whether keywords produce relevant images or not itself a difficult task.

While combining keywords from different systems, if a system perfoms good then keywords generated out of that system should be weighted more. Weightage for indiviudal system

In this system, we combine all the systems described above. We retrieve set of images for each system.

$$Weight = \frac{1}{\text{KL_Uni}} \tag{3.7}$$

where KL_Uni is KL-Divergence between unigram distribution of words in Article Text and Image Descriptions.

[1]Yahoo Content Analysis Platform APIs avialble at, https://developer.yahoo.com/search/content/V2/contentAnalysis.htr

| **Article** | **Image Description** |
|---|---|
| FORD IS to unveil a solar -powered concept car capable of 100 miles per gallon that offers the same performance as a plug-in hybrid , but without the need for a plug. The C-MAX Solar Energi Concept car, which will be revealed at this month's International CES gadget show in Las Vegas, uses a petrol engine combined with a gizmo that acts like a magnifying glass to concentrate the sun's rays on the vehicle's roof-mounted solar panels. Ford says the vehicle's estimated combined mileage is 100 miles per gallon. By using solar power instead of an electric plug, the manufacturer says a typical owner will reduce annual greenhouse gas emissions by four tonnes. The company sold about 85,000 hybrid or electric vehicles last year, including 6,300 units of its C-MAX Energi plug-in hybrid . | 1. A Ford C-MAX Hybrid is seen in a show room at the Ford Motor Co.'s Michigan Assembly Plant December 14, 2011 in Wayne, Michigan. Ford released details about the electrification of the Michigan Assembly Plant that will power production in part by one of the largest solar energy generator systems in order to produce their new C-MAX Hybrid and C-MAX Energi electric vehicles.<br>2. Workers build cars on the assembly line at the Ford Motor Co.'s Michigan Assembly Plant December 14, 2011 in Wayne, Michigan. Ford released details about the electrification of the Michigan Assembly Plant that will power production in part by one of the largest solar energy generator systems in order to produce their new C-MAX Hybrid and C-MAX Energi electric vehicles.<br>3. A worker installs a fuel tank on a Ford Focus on the assembly line at the Ford Motor Co.'s Michigan Assembly Plant December 14, 2011 in Wayne, Michigan. Ford released details about the electrification of the Michigan Assembly Plant that will power production in part by one of the largest solar energy generator systems in order to produce their new C-MAX Hybrid and C-MAX Energi electric vehicles.<br>4... |

Table 3.1: Article and descriptions of retrieved images after firing the query "**Solar!1000 AND Hybrid!950 AND (C-MAX Energi)!850**".

$$KL_Uni = \sum_{u \in Unigrams} \ln(\frac{P(u)}{Q(u)}) * P(u)$$

Here, P(u) is probablity of word u occuring in article. Q(u) si probablity of word u occuring in image descriptions.

### 3.4.6 Rank Aggregation-2/Overlapping Bigrams

It is observed that, in the previous we used distance between unigrams distribution. but not all words are important in the articles (like stopwords). Their presence in the articles and descriptions increase weight for the system.

$$Weight = \frac{|Bigrams(Article) \cap Bigrams(Des)|}{|Bigrams(Article)|} \tag{3.8}$$

### 3.4.7 Rank Aggregation-3/KL Divergence-Bigrams

In this method, we weight is calculated as linear combination of KL divergence on unigram distributions and bigram distributions of articles and image description.

$$Weight = \alpha * \frac{1}{KL_Bigram} + \beta * \frac{1}{KL_Uni} \tag{3.9}$$

$$KL_Uni = \sum_{u \in Bigrams} \ln(\frac{P(u)}{Q(u)}) * P(u)$$

Here, P(u) is probablity of bigram u occuring in article. Q(u) is probablity of bigram u occuring in image descriptions.

**Algorithm 1** Reranking algorithm

$Systems \Leftarrow \{TFIDF, RAKE, CapGen, Yahoo - CAP\}$

**for** System *sys* ∈ *Systems* **do**

  $Keywords_{sys}$ ⇐ Fecth Keywords of sys for the article

**end for**

keywords ⇐ {}

**for** System *sys* ∈ *Systems* **do**

  Query Q ⇐ Formulate Query from $Keyword_{sys}$

  ImageResults ⇐ Perform Image Search using Query Q.

  Weight ⇐ Calculate(Article, ImageDescription)

  keywords ⇐ keywords + Weight * $Keyword_{sys}$

**end for**

Query Q ⇐ Formulate Query from $Keywords$

ImageResults ⇐ Perform Image Search using Query Q.

Output a first Image

## 3.5 Summary

In this chapter, we described the experiments done in unsupevised and supervised settings. First unsupervised approach is based on number of occurences and proximity. Second unsupervised approach is based on modified textrank which includes co-referencing for constructing a text graph.

We also discussed about caption generation methods, and rank aggregation framwork for image recommendation.

# Chapter 4

# Experiments on Keyword Extraction and Image Retrieval

In this chapter, we present the experimentations performed on Caption Generation and Rank Aggregation models.

## 4.1 Experiments

| Type | Count |
|---|---|
| News | 50 |
| Entertainment | 46 |
| Cricket | 51 |
| Total | 147 |

Table 4.1: Corpus for manual evaluation

We have collected 20,000 articles from Yahoo news site. Each article is associated with one or more images with image description and title. Title contains on the average 8-10 words, on the other hand description is much more verbose which has around 50-70 words on average. Yahoo Image Search System [1] is meta data based search system which uses

[1]http://images.search.yahoo.com/images

| System | News | Entert. | Cricket | All |
|---|---|---|---|---|
| CapGen | 10.20 | 17.07 | – | – |
| TFIDF | **46.93** | 60.96 | 41.17 | 49.69 |
| TextRank | 14.28 | 09.75 | 01.96 | 08.66 |
| Yahoo-Cap | 38.77 | 58.54 | **45.09** | 48.28 |
| RR1 | 26.53 | 36.58 | 23.52 | 28.88 |
| RR2 | 36.73 | 60.97 | 39.21 | 45.64 |
| RR3 | 46.93 | **65.85** | **45.09** | **52.63** |

Table 4.2: Prescision@1 for relevancy of retrieved image

both title and description to fetch relevant set of images for a given query.

We use the image captions to generate ngram language model, which is used for reordering choosen words in Caption Generation Phase. SRILM toolkit is used for building language model.

For evaluating image recommendation relevancy, we manually **Manual Evaluation** For P@1 and P@5, 200 articles sampled from 20000 articles. Manually relevancy of recommended images evaluated.

**Automatic Evaluation** - To check the effectiveness of keywords extracted, we calculate percentage of keywords generated from our system overlaps with image caption.

## 4.2 Examples

In this section, we discuss the results generated from all systems.

In the example (Table 4.2 ), Caption generation fails to bring the rank entity 'Argentina'. It is due to similarity functions used in place of transition probabilty, the world similarity measures works well with nouns and verbs. The wordnet similarity functions failed to detect the similarity between named entities.

| System | News | Entert. | Cricket | All |
|---|---|---|---|---|
| CapGen | 12.92 | 09.32 | – | – |
| TFIDF | **48.97** | 43.13 | 30.71 | 40.93 |
| TextRank | 09.52 | 07.35 | 03.26 | 06.71 |
| Yahoo-Cap | 45.57 | **53.43** | 39.87 | 46.29 |
| RR1 | 17.68 | 35.25 | 22.88 | 25.27 |
| RR2 | 34.02 | 49.01 | 41.83 | 45.62 |
| RR3 | 44.22 | 50.49 | **46.41** | **47.04** |

Table 4.3: Prescision@5 for relevancy of retrieved image

| System | Keywords |
|---|---|
| CapGen | heres, closed, day, open, years, whats |
| TFIDF | heres, whats, closed, day, years, open |
| TextRank | **new years day**, whats open |
| RAKE | whats open, years day, closed, heres |
| Yahoo-Cap | heres, **new years day** |
| RR1 | **new years day**, **whats open**, heres, **closed** |
| RR2 | **new years day**, **whats open** |
| RR3 | **carrefour argentina**, **cakes**, **unusual ingredient**, **joked**,**cocaine**, **delicacy**, drug lord pablo escobar, erroneous labels, **cake supplier** |

Table 4.5: Example - Caption; whats open and closed on new years day

| **System** | **Keywords** |
|---|---|
| CapGen | **cakes**, **carrefour**, **ingredient**, ,joked,**cocaine**,**delicacy**,labels,**supplier**,lord,shelves |
| TFIDF | **carrefour**,**cakes**, **argentina**, joked, reassure, ingredient, **supplier**, **cocaine**, legalized, lilia |
| TextRank | company statement, brand cake, erroneous labels, many people, unusual ingredient, **cake supplier** |
| RAKE | social media user called lilia, neighboring uruguay legalized marijuana, cake supplier listed 12 grams, drug lord pablo escobar |
| Yahoo-Cap | **carrefour argentina**, **cakes**, the french supermarket chain, drug lord pablo escobar, delicacy, gustavo ramirez, stores |
| RR1 | the french supermarket chain, drug lord pablo escobar, **carrefour argentina**, **cake supplier**, unusual ingredient, **cakes**, gustavo ramirez, **joked**, **delicacy**, **reassure** |
| RR2 | **carrefour argentina**, **unusual ingredient**, **cakes**, company statement, erroneous labels, drug lord pablo escobar, **cake supplier**, **joked**, brand cake, **cocaine** |
| RR3 | **carrefour argentina**, **cakes**, **unusual ingredient**, **joked**,**cocaine**, **delicacy**, drug lord pablo escobar, erroneous labels, **cake supplier** |

Table 4.4: Example - Caption; carrefour argentina has scrambled to reassure customers after a delicacy from a cake supplier listed 12 grams of cocaine as an ingredient

## 4.3 Results

Table 4.3 compares the re-ranked results with the above baseline keyword extraction methods. The table clearly shows that images retrieved using re-ranked keywords as query, outperforms stated baseline algorithms. In all article types, TFIDF is performing close to the better performing system (Re-rank-4). However, Yahoo term extraction works better on the cricket articles because almost all news articles belongs to famous entities present in Open Data.., like Wikipedia.

| **System** | **Precision** (Keywords) |
|---|---|
| CapGen | 13.1440 |
| TFIDF | 16.1934 |
| TextRank | 09.7163 |
| RAKE | 09.7747 |
| Yahoo-Cap | 14.5922 |
| RR1-Our System | 16.2718 |
| RR2-Our System | 16.9454 |
| RR3-Our System | **17.5169** |

Table 4.6: Precision: Percentage of keywords generated matches with the words in the description of Images.

## 4.4 Summary

In this chapter, we discussed experimental setup, corpus, result, and examples for rank aggregation methods and caption generation methods. Next chapter, we discuss about submodularity.

# Chapter 5

# Submodularity and Summarization

The basic question that the summary has to answer are:

- How to select representative sentences?
- Which sentences should we select that best summarize(represent) a document?

[19] treated this task as monotone submodular function maximization. Further, they argued that monotone nondecreasing submodular functions are an ideal class of functions to investigate for document summarization. They show, in fact, that many well-established methods for summarization ([3], [7], [34]) correspond to submodular function optimization, a property not explicitly mentioned in these publications. They have taken this fact, however, as testament to the value of submodular functions for summarization: if summarization algorithms are repeatedly developed that, by chance, happen to be an instance of a submodular function optimization, this suggests that submodular functions are a natural fit.

## 5.1 Optimization Problem Formulations

Let the ground set V represents all the sentences (or other linguistic units) in a document (or document collection, in the multi-document summarization case). The task of extractive document summarization is to select a subset $S \in V$ to represent the entirety (ground set V ). There are typically constraints on S, however. Obviously, we should have $|S| < |V| = N$ as it is a summary and should be small. In standard summarization tasks (e.g., DUC evaluations), the summary is usually required to be length-limited. Therefore, constraints on S can naturally be modeled as knapsack constraints:

$$\sum_{i \in S} c_i < b \tag{5.1}$$

, where $c_i$ is the non-negative cost of selecting unit i (e.g., the number of words in the sentence) and b is our budget. If we use a set function $F : 2^V \to R$ to measure the quality of the summary set S, the summarization problem can then be formalized as the following combinatorial optimization problem:

$$S^* \in argmax_{S \subset V} F(S) \text{ subject to: } \sum_{i \in S} c_i < b \tag{5.2}$$

Note: In previous paper, the author et al. had considered cardinality constraint i.e. $c_i = 1$ and had proposed a graph based submodular selection.

The above problem is NP hard but a modified greedy algorithm with partial enumeration can solve above problem near-optimally with $(1 - e^{-1})$-approximation factor if F is monotone submodular [35].

Note: Sviridenko extended the proof given by Wolsey [39] for $c_i$ without cardinality constraint.

Another perspective is to treat the summarization problem as finding a low-cost subset of the document under the constraint that a summary should cover all (or a sufficient amount of) the information in the document, i.e.

$$S^* \in argmin_{S \subset V} \sum_{i \in S} c_i \text{ subject to:} F(S) > \alpha \tag{5.3}$$

However this type of framework does not satisfy a summary's budget constraint and thus, the author discarded such optimization problem.

Since, the greedy algorithm only gives optimal bound on the monotone submodular objectives; the author has proposed the following class of submodular functions:

$$F(S) = L(S) + \lambda R(S) \tag{5.4}$$

, where $F(S)$, total utility of summary is given as a linear combnation of $L(S)$, relevance and $R(S)$, non-redundancy.

Two properties of a good summary are relevance and non-redundancy. Objective functions for extractive summarization usually measure these two separately and then mix them together trading off encouraging relevance and penalizing redundancy. The redundancy penalty usually violates the monotonicity of the objective functions (Carbonell and Goldstein, 1998; Lin and Bilmes, 2010). The author therefore proposes to positively reward diversity instead of negatively penalizing redundancy. $L(S)$ measures the coverage, or âĂIJfidelityâĂİ, of summary set $S$ to the document, $R(S)$ rewards diversity in $S$, and $\lambda$ is a trade-off coefficient.

$$L(S) = \sum_{i \in V} \min\{c_i(S), \alpha c_i(V)\} \tag{5.5}$$

$$c_i(S) = \sum_{j \in S} w_{i,j} \tag{5.6}$$

Here, $w_{i,j} > 0$ measures the similarity between $i^{th}$ and $j^{th}$ sentences and thus, $c_i(S)$ measures the similarity of summary with the document.

$$R(S) = \sum_{i=1}^{K} \sqrt{\sum_{j \in P_i \cap S} r_j} \tag{5.7}$$

Here $P_i$; $i = 1 \ldots K$ is a partition of the ground set $V$ (i.e., $\cup_i P_i = V$ and the $P_i$s are disjoint) into separate clusters, and $ri > 0$ indicates the singleton reward of $i$ (i.e., the reward of adding $i$ into the empty set). The value $ri$ estimates the importance of $i$ to the summary. The function $R(S)$ rewards diversity in that there is usually more benefit to selecting a sentence from a cluster not yet having one of its elements already chosen. As soon as an element is selected from a cluster, other elements from the same cluster start having diminishing gain, because of the square root function. For instance, consider the case where $k_1 \in P_1$; $k_2 \in P_2$, $k_3 \in P3$, and $r(k_1) = 4$, $r(k_2) = 9$, and $r(k_3) = 4$. Assume $k_1$ is already in the summary set S. Greedily selecting the next element will choose $k_3$ rather than $k_2$ since $\sqrt{13} < 2 + 2$. In other words, adding $k3$ achieves a greater reward as it increases the diversity of the summary (by choosing from a different cluster).

Note: $R(S)$ is distinct from $L(S)$ in that $R(S)$ might wish to include certain outlier material that $L(S)$ could ignore.

$r_j$ is the average similarity of sentence $i$ to the rest of the document. It basically states

that the more similar to the whole document a sentence is, the more reward there will be by adding this sentence to an empty summary set. It is given by,

$$r_i = \frac{1}{N} \sum_j w_{i,j} \tag{5.8}$$

Note: $w_{i,j} > 0$ is the similarity between sentence $i$ and $j$ and thus, can be modeled as Jaccard or cosine similarity.

## 5.2 Proof of problem

The problem formulation can be proved as a monotone submodular function optimization.

Proof: To prove $F(S)$ is submodular, we can individually show that $L(S)$ and $R(S)$ are submodular and thus, by closedness property, $F(S)$ is submodular.

Theorem: Given functions $F : 2^V \to R$ and $f : R \to R$, the composition $F^* = foF : 2^V \to R$ (i.e., $F^*(S) = f(F(S))$) is non-decreasing submodular, if f is non-decreasing concave and F is non-decreasing submodular.

Proof: (not given in paper) Intutively in special cases, it can be proved by the Submodularity and Concavity theorem, according to which if there exists $g : N \to R$ and $F(A) = g(|A|)$ Then, F(A) is submodular iff g is concave. Composition of two concave functions is concave and composition of two non-decreasing functions is non-decreasing. Therefore, if there exists $h : N \to R$ and $F^*(A) = h(|A|)$ and $h = fog$ and h is non-decreasing concave, thus $F^* = foF$ is non-decreasing submodular.

Claim: $L(S)$ is monontone submodular.

Proof: Clearly, $c_i(S)$ is a monotone non- decreasing submodular function as, $c_i(A \cup k) - c_i(A) \geqslant c_i(A \cup j \cup k) - c_i(A \cup j)$ since, $c_i(A \cup k) - c_i(A) = w_{i,k}$ and $c_i(A \cup j \cup k) - c_i(A \cup j) = w_{i,k}$

Adding a new sentence to any set will add the similarity of the new sentence with document, irrespective of set size. Consider the fact that $f(x) = min(x, a)$ where $a > 0$ is a concave monotone non-decreasing function, and by Theorem based on composition of monotone non-decreasing concave function with monotone non- decreasing submodular function (given in paper), each summand in $L(S)$ is a monotone non-decreasing submodular function, and as summation preserves submodularity by closedness property, $L(S)$ is sub-modular. $L(S)$ is also monotone non-decreasing as summation of monotone non-decreasing function is monotone non-decreasing. Hence, proved.

Claim: $R(S)$ is monontone submodular.

Proof: $R(S)$ is submodular by using the composition rule from above theorem. The square root is non-decreasing concave function. Inside each square root lies a modular function with non-negative weights (and thus is monotone). Applying the square root to such a monotone submodular function yields a submodular function, and summing them all together retains submodularity, as summation preserves submodularity by closedness property. The monotonicity of $R(S)$ is straightforward.

The functions $L(S)$ can also be modelled as:

1. $L(S) = \sum_{i \in V, j \in S} w_{i,j}$ (modular objective)

2. $L(S) = \sum_{i \in V} max_{j \in S} w_{i,j}$ (facility location objective)

3. $L(S) = \sum_{i \in \tau(S)} c_i$ (concept based objective)

Similarly, for $R(S)$,

1. Instead of using a ground set partition, intersecting clusters can be used.

2. The square root function can be replaced with any other non-decreasing concave functions (e.g., $f(x) = log(1 + x)$) while preserving the desired property of $R(S)$, and the curvature of the concave function then determines the rate that the reward diminishes.

3. Multi-resolution clustering (or partitions) with different sizes (K) can be used, i.e., we can use a mixture of components, each of which has the structure of $R(S)$ Eqn.

Each of the above formulations are monotone sub-modular.

## 5.3 Optimization Algorithm

In [17], the author propose algorithm 5.1 to solve Eqn. (3.2). The algorithm sequentially finds unit k with the largest ratio of objective function gain to scaled cost, i.e., $\frac{f(G \cup \{l\}) - f(G)}{(c_r)^l}$

, where $r > 0$ is the scaling factor. If adding k increases the objective function value while not violating the budget constraint, it is then selected and otherwise bypassed. After the sequential selection, set G is compared to the within-budget singleton with the largest objective value, and the larger of the two becomes the final output.

**Algorithm 1** Modified greedy algorithm

$G \leftarrow \emptyset$
$U \leftarrow V$
**while** $U \neq \emptyset$ **do**
  $k \leftarrow \arg\max_{\ell \in U} \frac{f(G \cup \{\ell\}) - f(G)}{(c_\ell)^r}$
  $G \leftarrow G \cup \{k\}$ **if** $\sum_{i \in G} c_i + c_k \leq B$ **and** $f(G \cup \{k\}) - f(G) \geq 0$
  $U \leftarrow U \setminus \{k\}$
**end while**
$v^* \leftarrow \arg\max_{v \in V, c_v \leq B} f(\{v\})$
**return** $G_f = \arg\max_{S \in \{\{v^*\}, G\}} f(S)$

Figure 5.1: Modified Greedy Algorithm[1]

The algorithm sequentially finds unit k with the largest ratio of objective function gain to scaled cost, i.e., $\frac{(f(G \cup \{l\}) - f(G))}{c_l^r}$, where $r > 0$ is the scaling factor. If adding k increases the objective function value while not violating the budget constraint, it is then selected and otherwise bypassed. After the sequential selection, set G is compared to the within-budget singleton with the largest objective value, and the larger of the two becomes the final output. The essential aspect of a greedy algorithm is the design of the greedy heuristic. As discussed in (Khuller et al., 1999), a heuristic that greedily selects the k that maximizes $\frac{(f(G \cup \{l\}) - f(G))}{c^k}$ has an unbounded approximation factor. For example, let $V = a, b, f(a) = 1, f(b) = p, c_a = 1, c_b = p + 1, and B = p + 1$. The solution obtained by the greedy heuristic is $a$ with objective function value 1 , while the true optimal objective function value is p. The approximation factor for this example is then p and therefore unbounded. We address this issue by the following two modifications to the naive greedy algorithms. The first one is the final step (line 8 and 9) in Algorithm 1 where set G and singletons are compared. This step ensures that we could obtain a constant approximation factor for r = 1. The second modification is that we introduce a scaling factor r to adjust the scale of the cost. Sup- pose, in the above example, we scale the cost as $c_a = 1^r, c_b = (p+1)^r$, then selecting a or b depends also on the scale $r$, and we might get the optimal solution using a appropriate $r$.

## 5.4 Results

Their implemenation had better than existing state-of-the-art performance on a number of standard summarization evaluation tasks, namely DUC-04 through to DUC-07. and their system got good result with 39.35% Recall, F-score 38.90%.

# Chapter 6

# Submodularity and Opinion Summarization

Opinion Summarization can also be modeled as a submodular optimization problem, since the subjectivity also holds a diminishing return property like summary. Many researchers, who modelled the opinion summarization problem as an objective function used submodular function unkowingly like [15] and [30]. In this section, we will arue through examples that how opinion summarization can be modelled as submoduar objective.

## 6.1 Relation to Previous Works

Previous works in opinion summarization has already been discussed in Chapter 2. However, in this section, the previous work will be critiqued based on the nature of the optimization function choosen.

In [30] , a mincut-based algorithm was proposed to classify each sentence as being subjective or objective. The purpose of this work was to remove objective sentences from reviews to improve document level sentiment classification. Interestingly, the cut functions are symmetrical and submodular, and the problem of finding min-cut is equivalent to minimizing a symmetric submodular function.

[15] proposed three different models - sentiment match (SM), sentiment match + aspect coverage (SMAC) and sentiment-aspect match (SAM) to perform summarization of reviews of a product. The first model is called sentiment match (SM), which extracts sen-

tences so that the average sentiment of the summary is as close as possible to the average sentiment rating of reviews of the entity *i.e.* low MISMATCH but with high sentiment INTENSITY. The second model, called sentiment match + aspect coverage (SMAC), builds a summary that trades-off between DIVERSITY, maximally covering important aspects and MISMATCH, matching the overall sentiment of the entity along with high INTENSITY. The third model, called sentiment-aspect match (SAM), not only attempts to cover important aspects, but cover them with appropriate sentiment using KL-Divergence function. Here, INTENSITY and DIVERSITY in the first two models are linear monotone submodular functions, while KL-Divergence function *i.e.* relative entropy in last model, unlike entropy is not monotone submodular.

On the other hand, [19] treated the task of generic summarization as monotone submodular function maximization. Further, they argued that monotone non-decreasing submodular functions are an ideal class of functions to investigate for document summarization. They also show, in fact, that many well-established methods for summarization ([3], [7], [34]) correspond to submodular function optimization, a property not explicitly mentioned in these publications. Since many authors either in summarization or opinion summarization have used functions similar to submodular functions as objective, we can take this fact as testament to the value of submodular functions for opinion summarization.

## 6.2 Background on Submodular Functions

A submodular function is a set function ($f : 2^V \rightarrow R$) having a natural diminishing returns property. Diminishing returns property holds if the difference in the value of the function that a single element makes when added to an input set decreases as the size of the input set increases *i.e.* for every $A, B \subseteq V$ with $A \subseteq B$ and every $x \in V \backslash B$, we have that $f(A \cup \{x\}) - f(A) \geqslant f(B \cup \{x\}) - f(B)$. A submodular function f is monotone if for every $A \subseteq B$, we have that $f(A) \leqslant f(B)$.

The extractive summarization can be modeled as optimization problem *i.e.* finding a set $S \subseteq V$ (S is set of sentences in summary, V is set of sentences in Document) which maximizes a submodular function $f(S)$ subject to budget constraints. In the following section, we will justify the use of submodular function for opinion summarization. Another ad-vantage of choosing monotone submodular function is that there exists a polynomial-time greedy algorithm for constrained monotone submodular objective. The greedy algorithm guarantees that the summary solution obtained is almost as good as (63%) the best possible summary solution according to the objective [39; 35].

For more information on monotone submodular functions, please refer appendix.

## 6.3 Submodularity in Opinion Summarization

Opinion Summarization should be modeled as a monotone submodular optimization problem, since opinion summary also holds following properties:

- Monotonicity - As more sentences are added to opinion summary, subjectivity increases along with information content as opinionated words are being added.
- Diminishing Return - If multiple sentences of varying intensity are added to opinion summary, the effect of lower intensity dilutes in presence of higher intensity bearing polar sentence in the summary.

### 6.3.1 Examples from Cricket Domain

Consider following sentence of positve polarity:
A: *Rahul Dravid is a great batsman.*
Consider another sentence with same polarity:
B: *Rahul Dravid is a very consistent player.*
Now, consider following text, which conatins both of the above sentence:
$(A \cup B)$: *Rahul Dravid is a great batsman. He is a very consistent player.*

Consider the difference between singleton set B and its superset, $(A \cup B)$. After reading the text, it seems clear than the effect of sentence B has diminished in front of sentence A in text $(A \cup B)$, though both are of same polarity.

The diminishing return not only holds for same polarity but also, for opposite polarity.

Consider following sentence of positve polarity:
A: *Dhoni is a great batsman.*
Consider following sentence of negativee polarity:

B: *Dhoni's wicket keeping skills are not upto Dravid's.*
Now, consider following text, which conatins both of the above sentence:
$(A \cup B)$: *Dhoni is a great batsman. His wicket keeping skills are not upto Dravid's.*
After reading the text, it seems clear than the effect of sentence B has diminished in front of sentence A in text $(A \cup B)$, as usually polarity of higher intensity dominates over the polarity of lower intensity.

Now, lets consider a general example,

"*Sachin is a great batsman. His backfoot batting is unmatchable. He also bowls decent spin.*"

If the budget had been only two subjective sentences, then picking up first two would have redudantly captured only single aspect of Sachin(i.e. batting) and the redundancy of the concept(batting) causes a diminishing return from the second sentence. However, picking the last sentence with the one of the first two would have covered both aspects(i.e. batting and bowling) and since, the sentences are not overlapping in aspects, there is no diminishing return.

### 6.3.2 Examples from Movie Domain

To show that opinion summarization inherently follow the diminishing return property. Lets consider the following sentences with positive polarity:
**A**: *The acting in Hell is also excellent, with the dreamy Depp turning in a typically strong performance and deftly handling a british accent .*
**B**: *Worth mentioning are the supporting roles by Ians Holm and Richardsonlog.*
**(A ∪ B)** : *The acting in Hell is also excellent, with the dreamy Depp turning in a typically strong performance and deftly handling a british accent . Worth mentioning are the supporting roles by Ians Holm and Richardsonlog.*
Consider the difference between singleton set B and its superset, $(A \cup B)$. Sentence A and B convey positive sentiment, but sentence B has less intensity compared to sentence A. After reading the text (A ∪ B), it is clear that the effect of sentence B has diminished in front of sentence A, though both are of same polarity. The diminishing return not only holds for same polarity but also, for opposite polarity. Lets consider another example:

**A**: *The movie is predictive with foreseeable ending.*
**B**: *Still it's very well-done that no movie in this entire year has a scene that evokes pure joy as this does.*
of the above sentence:
**(A ∪ B)** : *The movie is predictive with foreseeable ending. Still it's very well-done that no movie in this entire year has a scene that evokes pure joy as this does.*
Sentence A has negative sentiment whereas sentence B conveys positive sentiment with more intensity. When we read the text (A ∪ B), it is clear that the effect of sentence A has diminished in front of sentence B in text , as usually polarity of higher intensity dominates over the polarity of lower intensity.

Now, lets consider a general example, "*Laurence plays Neo's mentor Morpheus and he does an excellent job of it. His lines flow with confidence and style that makes his acting unique and interesting. The movie has lot of special effects and action-packed scenes with part of the appeal has philosophical and religious underpinnings.*" If the budget had been only two subjective sentences, then picking up first two would have redundantly captured only single aspect (*i.e.* acting) and the redundancy of the concept (acting) also causes a diminishing return of the second sentence because of the difference in sentiment intensity. However, picking the last sentence with either one of the first two would have not just covered both the aspects (*i.e.* acting and visual effects) but since, the sentences are not overlapping in aspects, there would not have been any diminishing return of sentiment on shared aspect (acting).

Thus, it can be verified that opinion polarity also holds submodular property of diminishing return, if they are on the same aspect of a distinct entity. So, we need to model this submodular proprty of opinion through relevant submodular functions.

## 6.4 Optimization Problem Formulations

We propose a submodular objective for opinion summarization similar to summarization as proposed in [19].
Let the ground set V represents all the sentences (or other linguistic units) in a document (or document collection, in the multi-document summarization case). The task of extractive opinion summarization is to select a subset $S \in V$ to represent the entirety (ground set $V$). There are typically constraints on S, however. Obviously, we should have $|S| < |V| = N$ as it is a summary and should be small. Therefore, constraints on S can naturally be modeled as knapsack constraints:

$$\sum_{i \in S} c_i < b \tag{6.1}$$

, where $c_i$ is the non-negative cost of selecting unit i (e.g., the number of words in the sentence) and b is our budget. If we use a set function $F : 2^V \to R$ to measure the quality of the summary set S, the summarization problem can then be formalized as the following combinatorial optimization problem:

$$S^* \in argmax_{S \subset V} F(S) \text{ subject to: } \sum_{i \in S} c_i < b \tag{6.2}$$

The total utility of summary, $F(S)$ will be given as a linear combnation of $L(S)$, relevance and $R(S)$, non-redundancy of aspect coverage and $B(S)$, subjectivty score.

$$F(S) = \alpha L(S) + \beta R(S) + \gamma B(S) \tag{6.3}$$

However, the aspect coverage and subjectivity score can be combined to formulate as:

$$F(S) = \alpha L(S) + \beta A(S) \tag{6.4}$$

, where $F(S)$, total utility of summary is given as a linear combnation of $L(S)$, relevance and $A(S)$, subjective coverage of aspects.

This formulation clearly brings out the trade-off between the subjective and the objective part. The intuition behind the combination of sentiment and aspect coverage in same function A(S) is that opinion polarity holds submodular property of diminishing return only if the set of sentences talk about common aspect of the same entity as discussed in previous section. L(S) , relevance is modeled same as in Bilmes et al. as it captures the summary property, while A(S) has been modeled differently through a suitable submodular function such that it captures the subjectivity property.

$$L(S) = \sum_{i \in V} min\{c_i(S), \gamma c_i(V)\} \tag{6.5}$$

$$c_i(S) = \sum_{j \in S} w_{i,j} \tag{6.6}$$

Here, $w_{i,j} > 0$ measures the similarity between $i^{th}$ and $j^{th}$ sentences and $c_i(S)$ measures the similarity of summary with the document.

## 6.5 Possible Submodular Subjectivity Score

We can use the following scores in the submodular formulation to quantify the subjective aspect of the word:

- Probability Score of sentences
- Unsupervised Score of each word - Senti-Wordnet Score
- Supervised Score of each word: Chi-Square Score (learned from training corpus)

However, Probability scores are not appropriate functions to quantify submodular property of sentences as they are not submodular functions. Probabilty Scores are bounded between 0 and 1 and since, a function canâĂŹt be all - bounded, monotone and submodular at the same time, they are not submodular. Probabilty values are evaluated for set of sentence, which are negatively penalising(strict no for a submoular function - [17].

Consider the following counter-example:

$A$: *This hotel is nice.*

$P(A) = .8$

$A \cup B$: *This hotel is nice. It has around 100 staffs and a large banquet hall. It has many rooms too.*

$P(A \cup B) = .7$ (due to addition of objective words)

In the above example, the probability scores violates monotonicity.

## 6.6 Proposed Monotone Submodular Formulations

Since, A(S) , subjective coverage of aspects has to be modeled as monotone submodular function, it has been formulated as :

- $A_1$. **Modular Function**

  $A_1$(S) is simple linear function, which is sum of weighted subjective scores for each sentence. No budgeting constraints are added to this formulation. Formulation similar to this is proposed in other opinion summarization literatures as this is the most intutive formulation.

$$A_1(S) = \sum_i \sum_{j \in (P_i \cap S)} s_j * w_i \tag{6.7}$$

**Algorithm 2** Calculate $A_1$(S) - Modular Function

$A(S) \Leftarrow 0.0$
**for** Aspect asp $\in$ Aspects **do**
  $aspScore[asp] \Leftarrow 0.0$
**end for**
**for** Sentence s $\in$ Summary **do**
  aspID $\Leftarrow$ Aspect or cluster of s.
  $newScore \Leftarrow aspScore[aspID] + subjectivityScore(s)$
  $aspScore[aspID] \Leftarrow newScore$
**end for**
**for** Aspect asp $\in$ Aspects **do**
  $A(S) \Leftarrow A(S) + aspScore[aspID]$
**end for**

- $A_2$. **Budget-additive Function**

  $A_2(S)$ is an extension to $A_1(S)$, where maximum subjectivity score is restricted with budget based on aspect. Here, $0 < \lambda < 1$ is threshold coefficient for budget additive function to avoid redundancy of high sentiment on same aspect. When aspect i is saturated by S ( $min(\sum_{j \in (P_i \cap S)} s_j, \lambda) = \lambda$), any new sentence j cannot further improve coverage over i and thus, other aspects, which are not yet saturated, will have a better chance of being covered. Thus, this formulation ensures that produced summary is diverse enough and conveys sentiment about different aspects by budgeting.

**Algorithm 3** Calculate $A_2(S)$ - Budget-additive Function

$A(S) \Leftarrow 0.0$
**for** Aspect asp $\in$ Aspects **do**
  $aspScore[asp] \Leftarrow 0.0$
**end for**
**for** Sentence s $\in$ Summary **do**
  aspID $\Leftarrow$ Aspect or cluster of s.
  $newScore \Leftarrow aspScore[aspID] + subjectivityScore(s)$
  $aspScore[aspID] \Leftarrow newScore$
**end for**
**for** Aspect asp $\in$ Aspects **do**
  $Score \Leftarrow min(aspScore[asp], Budget[asp])$
  $A(S) \Leftarrow A(S) + Score$
**end for**

$$A_2(S) = \sum_i min(\sum_{j \in (P_i \cap S)} s_j, \lambda) * w_i \tag{6.8}$$

- $A_3$. **Polarity Partitioned Budget-additive Function**

  In previous formulation we have not considered the polarity of the sentences. For example, if an aspect have many positive sentences with more intensity but few negative sentences with less intensity, $A_2$ more likely to reward more positive sentences because of intensity. In this formulation budgeting applied not only on aspect but polarity scores too. This ensures that both positive and negative polarity sentences are part of summary.

$$A_3(S) = \sum_i min(\sum_{j \in (P_i \cap S \cap pos)} s_j, \lambda) * w_i + min(\sum_{j \in (P_i \cap S \cap neg)} s_j, \lambda) * w_i \tag{6.9}$$

- $A_4$. **Facility Location Function**

**Algorithm 4** Calculate $A_3(S)$ - Polarity Partitioned Budget-additive Function

$A(S) \Leftarrow 0.0$
**for** Aspect asp $\in$ Aspects **do**
    $aspScorePos[asp] \Leftarrow 0.0$
    $aspScoreNeg[asp] \Leftarrow 0.0$
**end for**
**for** Sentence s $\in$ Summary **do**
    aspID $\Leftarrow$ Aspect or cluster of s.
    **if** $polarityScore(s) \leqslant 0.0$ **then**
        $newScore \Leftarrow aspScoreNeg[aspID] + subjectivityScore(s)$
        $aspScoreNeg[aspID] \Leftarrow newScore$
    **else**
        $newScore \Leftarrow aspScorePos[aspID] + subjectivityScore(s)$
        $aspScorePos[aspID] \Leftarrow newScore$
    **end if**
**end for**
**for** Aspect asp $\in$ Aspects **do**
    $NegScore \Leftarrow min(aspScoreNeg[asp], Budget[asp])$
    $PosScore \Leftarrow min(aspScorePos[asp], Budget[asp])$
    $A(S) \Leftarrow A(S) + NegScore + PosScore$
**end for**

Facility Location objective is monotone submodular introduced in Krause and Golovin, described, "Suppose we wish to select, out of a set V, some locations to open up facilities in order to serve a collection of m customers. If we open up a facility at location j, then it provides service of value $M_{ij}$ to customer i. If each customer chooses the facility with highest value, the total value provided to all customers is modeled by the set function."

Here we use the facility location function for opinion summarization as choosing sentences (locations) to serve collection of aspects (customers). So $A_4$ rewards only a sentence which has maximum subjectivity score in each aspect.

$$A_4(S) = \sum_i max_{j \in (P_i \cap S)} s_j \tag{6.10}$$

**Algorithm 5** Calculate $A_4(S)$ - Facility Location Function

```
A(S) ⇐ 0.0
for Aspect asp ∈ Aspects do
    aspScore[asp] ⇐ 0.0
end for
for Sentence s ∈ Summary do
    aspID ⇐ Aspect or cluster of s.
    newScore ⇐ max(aspScore[aspID], subjectivityScore(s))
    aspScore[aspID] ⇐ newScore
end for
for Aspect asp ∈ Aspects do
    A(S) ⇐ A(S) + aspScore[aspID]
end for
```

- $A_5$. **Polarity Partitioned Facility Location Function**

  $A_5$ is similar to $A_4$, but for each aspect $A_5$ rewards two sentences with positive and negative polarity and maximum subjectivity score.

$$A_5(S) = \sum_i max_{j \in (P_i \cap S \cap pos)} s_j + \sum_i max_{j \in (P_i \cap S \cap neg)} s_j \quad (6.11)$$

**Algorithm 6** Calculate $A_5(S)$ - Polarity Partitioned Facility Location Function

$A(S) \Leftarrow 0.0$

**for** Aspect asp $\in$ Aspects **do**

  $aspScorePos[asp] \Leftarrow 0.0$

  $aspScoreNeg[asp] \Leftarrow 0.0$

**end for**

**for** Sentence s $\in$ Summary **do**

  aspID $\Leftarrow$ Aspect or cluster of s.

  **if** $polarityScore(s) \leqslant 0.0$ **then**

    $newScore \Leftarrow max(aspScoreNeg[aspID], subjectivityScore(s))$

    $aspScoreNeg[aspID] \Leftarrow newScore$

  **else**

    $newScore \Leftarrow max(aspScorePos[aspID, subjectivityScore(s))$

    $aspScorePos[aspID] \Leftarrow newScore$

  **end if**

**end for**

**for** Aspect asp $\in$ Aspects **do**

  $A(S) \Leftarrow A(S) + \text{NegScore} + \text{PosScore}$

**end for**

Here $P_i; i = 1...K$ is a partition of the ground set $V$ (*i.e.*, $\cup_i P_i = V$), which contains sentences pertaining to different distinct aspects. $w_i$ are the weights of the partitions, based on the corresponding aspects. $s_j$ is the subjective score of the sentence j in summary. The subjective score $s_j$ is calculated using senti-wordnet as sum of the positive $\in [0, 1]$ and negative score $\in [0, 1]$ [5] .

$$s_j = \sum_{word \in sentence_j} sentiwordnet_pos_score(word) + sentiwordnet_neg_score(word) \tag{6.12}$$

$pos$ and $neg$ are the partition of the sentences in the ground set $V$, based on their sign of polarity score. The polarity score for partitioning sentences into $pos$ and $neg$ is calculated as difference of the positive and negative score.

$$polarityScore(j) = \sum_{word \in sentence_j} (sentiwordnet_pos_score(word) - sentiwordnet_neg_score( \tag{6.13}$$

Here, $0 < \lambda < 1$ is threshold coefficient for budget additive function to avoid redundancy of high sentiment on same aspect. When aspect i is saturated by S ( $min(\sum_{j \in (P_i \cap S)} s_j, \lambda) = \lambda$), any new sentence j cannot further improve coverage over i and thus, other aspects, which are not yet saturated will have a better chance of being covered. Polarity based partitions bring out contrast view on a particular aspect, which is similar to contrast view opinion summarization to give the reader a direct comparative view of different strong opinions [11].

Each of the above functions are monotone submodular as the parameters $s_j$ and $w_i$ are positive. Since the first function is linear, it is both submodular and supermodular, thus modular. Budget additive and facility location functions [12] are special types of monotone submodular functions. Since, monotone submodularity is preserved under non-negative linear combinations, polarity based partitioned function, whose sub-parts are monotone submodular is monotone submodular.

**Algorithm 7** Overall Algorithm - Summary Extraction

**for** *Sentence s* $\in$ *Document* D **do**

  *Assign sentence s to one of aspects in movie ontology.*

**end for**

$S \Leftarrow \{\}$

$budget \Leftarrow 200$

**while** *change in summary* S **do**

  *maxReturn* $\Leftarrow$ *0.0*

  *newSentence* $\Leftarrow$ *None*

  **for** *Sentence s* $\in$ *Document* D and *s* $\notin$ *summary* S **do**

    **if** $len(s) \geqslant budget$ **then**

      *continue*

    **end if**

    $S^* \Leftarrow S + \{s\}$

    $F(S^*) \Leftarrow \alpha L(S^*) + (1 - \alpha)A(S^*)$

    $return \Leftarrow \frac{F(S^*) - F(S)}{len(s)^r}$

    **if** $return \geqslant maxReturn$ **then**

      $maxReturn \Leftarrow return$

      $newSentence \Leftarrow s$

    **end if**

  **end for**

  **if** $newSentence \neq None$ **then**

    $S \Leftarrow S + newSentence$

    $budget \Leftarrow budget - len(newSentence)$

  **end if**

**end while**

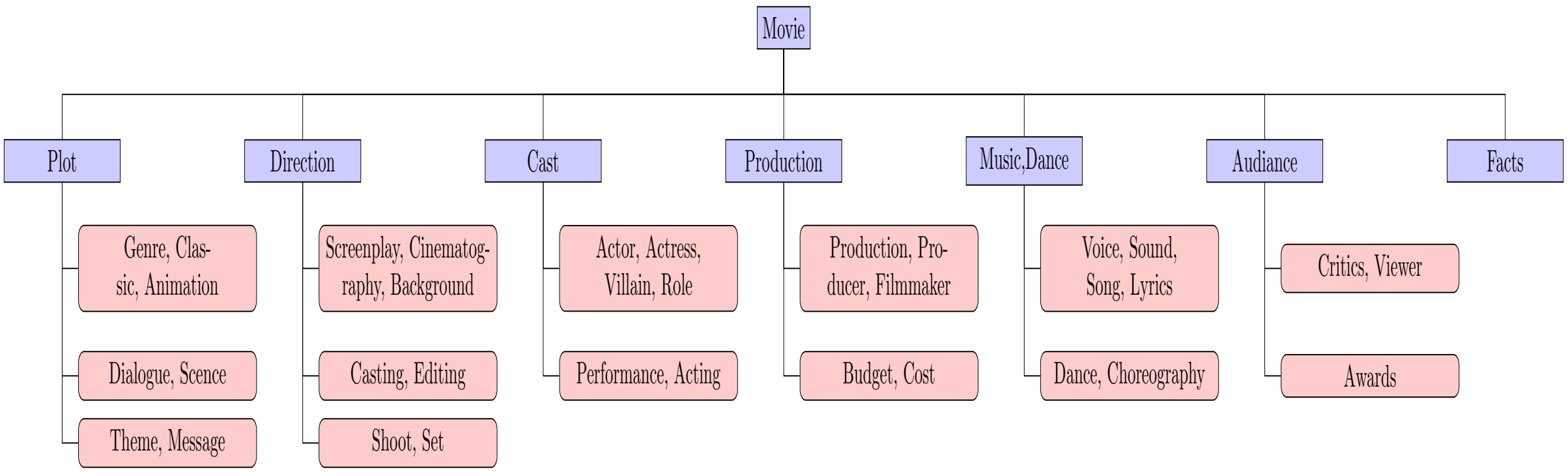

Figure 6.1: Movie Ontology Tree

## 6.7 Experiment

In this paper, we have selected movie domain documents for opinion summarization task as movie reviews have the following two important parts:

1. Plot - Description of the story in reviewer's words, which is factual in nature
2. Critique - Personal Opinion about the movie, which is sentiment bearing

Clearly, opinion summary to be generated will have trade-off between the two opposing parts - subjective critique and objective plot, on which we have striked out a balance through linear combination of suitable submodular functions in our paper.

We have used the polarity dataset from [30]. The dataset, containing 1000 positive and 1000 negative movie reviews has been divided into training and testing data. The testing data contains first 25 positive and 25 negative reviews from the dataset, while the remaining dataset is used for training as DUC benchmarks also have around 20 documents for evaluation.

Movie ontology tree (figure 6.1) is obtained from the training data using the ontology extractor based on k-partite graph learning algorithm [2]. The ontology is further enriched by adding corresponding clue words using wordnet sense propagation algorithm [5] for three iterations.

In the experiment, the modified greedy algorithm [17] is used for summary generation of 50 test documents within budget of 200 words. The basic textual/linguistic units we consider in our experiments are sentences. For each document cluster, sentences in all the documents of this cluster forms the ground set V. In the algorithm, the sentences are clustered in different partitions, corresponding to different nodes in the first two levels of the ontology tree using the clue words. The weights of the partitions as well as the threshold parameters for the A(S) are currently kept proportional to the inverse of the depth of that aspect in the ontology-tree as sentiment expressed on the concepts at higher level in the ontology tree should have more weightage.

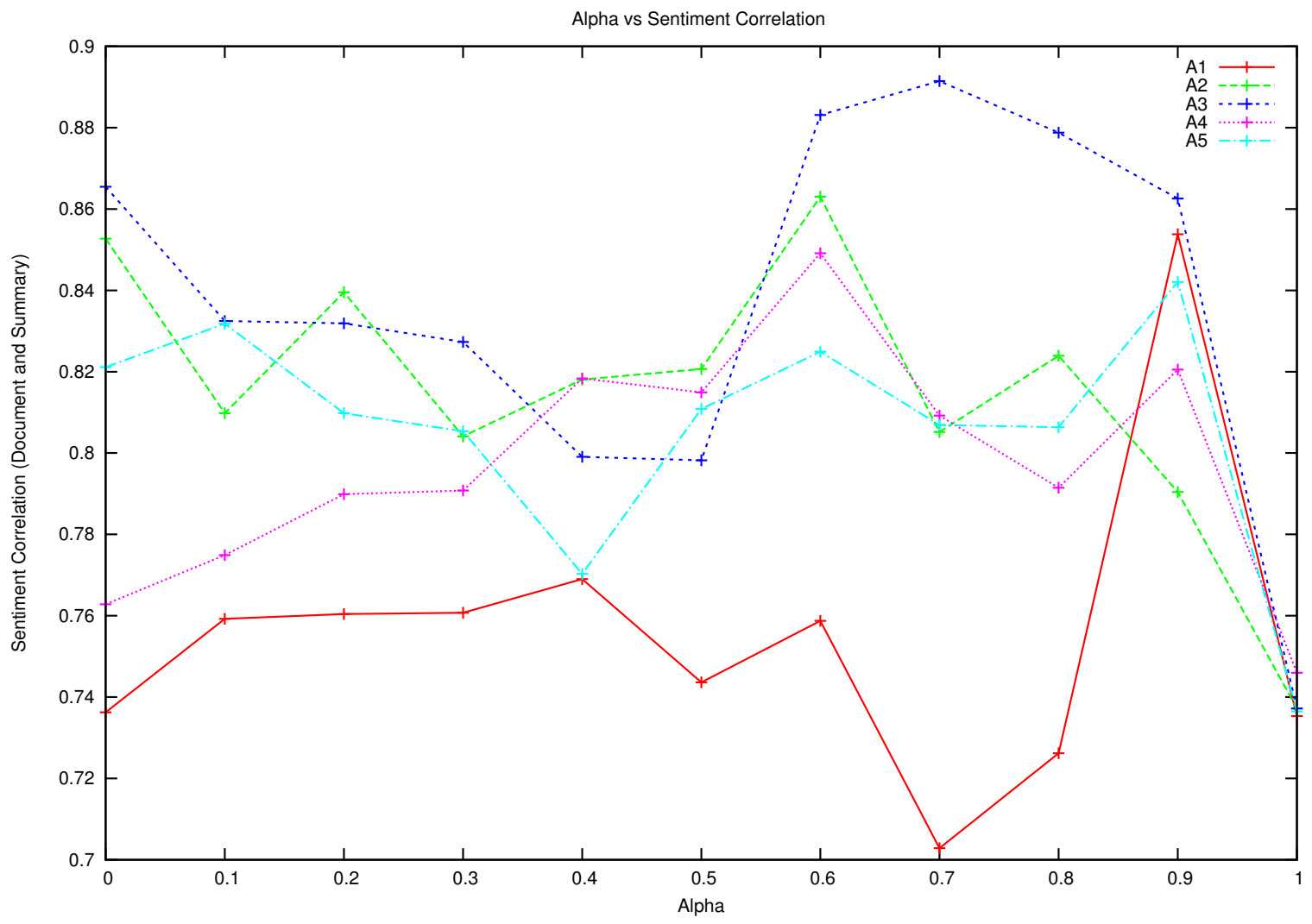

Figure 6.2: Sentiment Correlation *vs* $\alpha$ (r=0)

The linear combination parameter $\beta$ is set as $1 - \alpha$ to bring out the trade-off between relevance and subjective coverage of aspects and $\alpha$ is varied from 0 to 1. $r$ is varied from 0 to 1 in step size of 0.25 to observe the effect of scaling parameter on the summary. Graphs are plotted for comparing the five functions for different values of $\alpha$ and $r$. The expected shape of graph should be increasing for ROUGE score and decreasing for sentiment correlation for increasing value of $\alpha$ as $\alpha$ is proportional coefficient for relevance while $1 - \alpha$ is proportional coefficient for subjectivity.

We use ROUGE [16] for evaluating the content of summaries. We have used the 200 test documents that are manually summarized as gold standard data for ROUGE evaluation. For figuring out the sentiment correlation between manual and system generated summaries, we trained Naive Bayes sentiment classifier [33] on training data using bag of words approach with features as unigrams and bigrams and then, using minimum jaccard score of three for feature extraction for calculating the sentiment. The measure of sentiment preser-

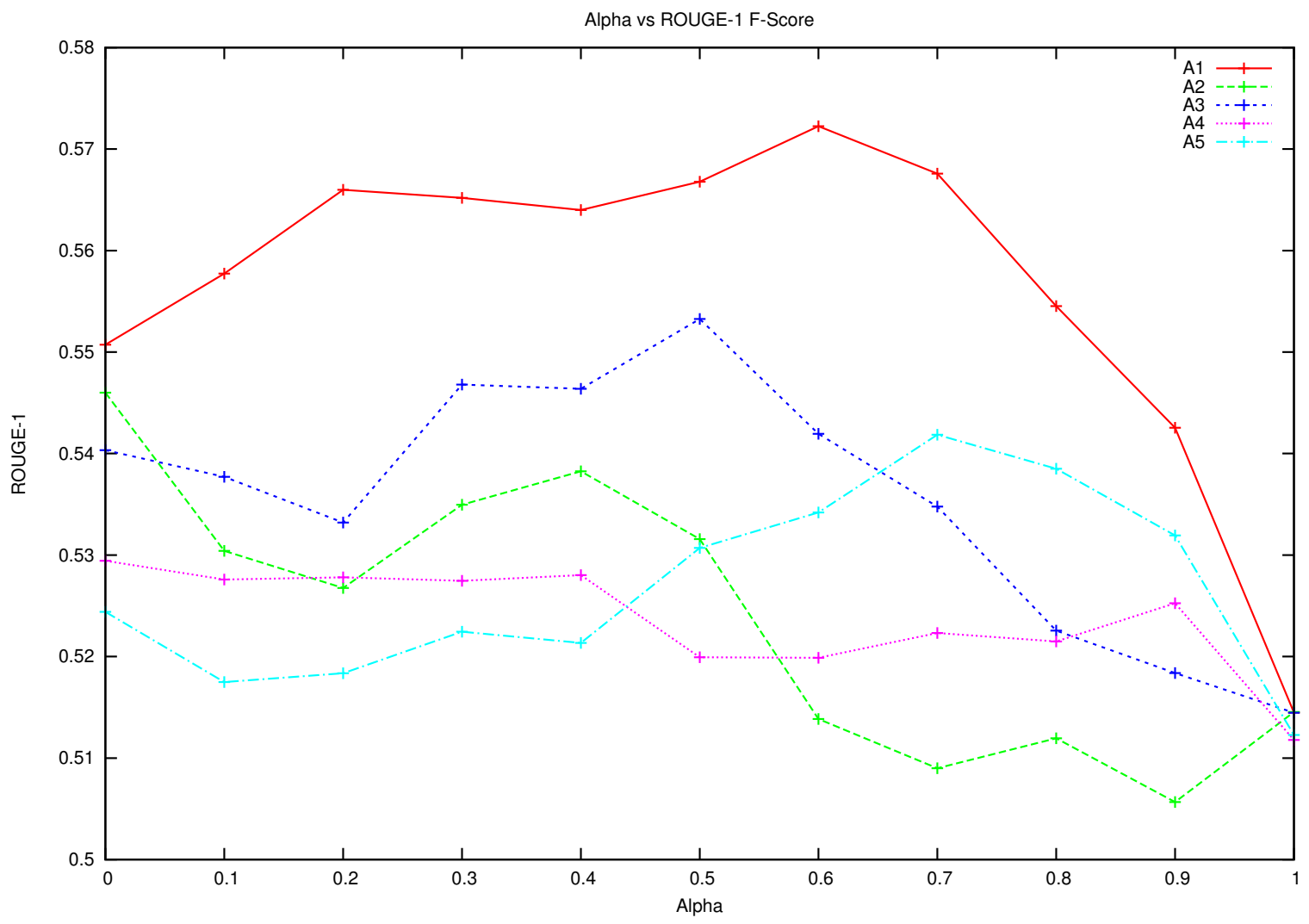

Figure 6.3: ROUGE 1 F-score *vs* $\alpha$ (r=0)

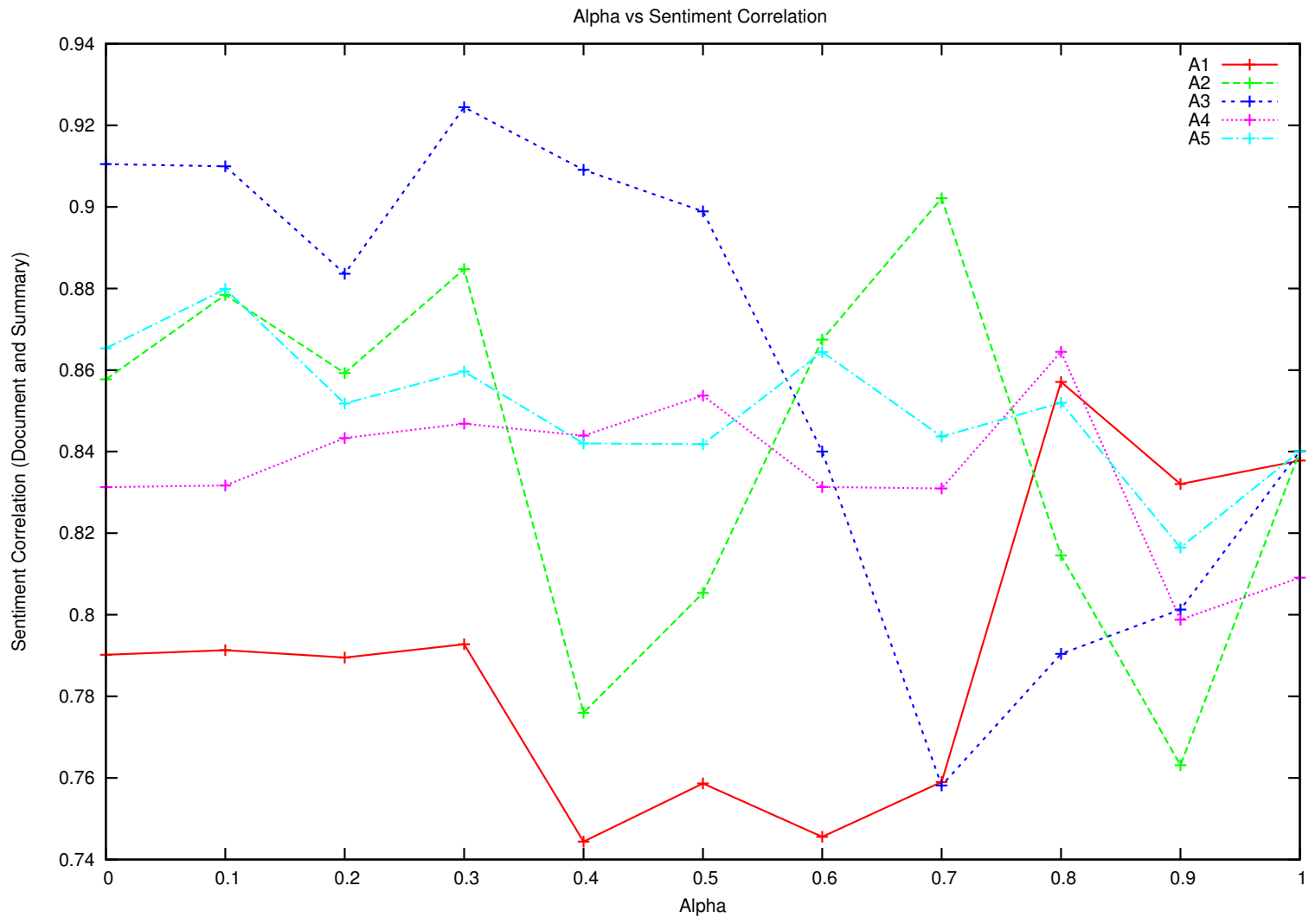

Figure 6.4: Sentiment Correlation *vs* $\alpha$ (r=0.25)

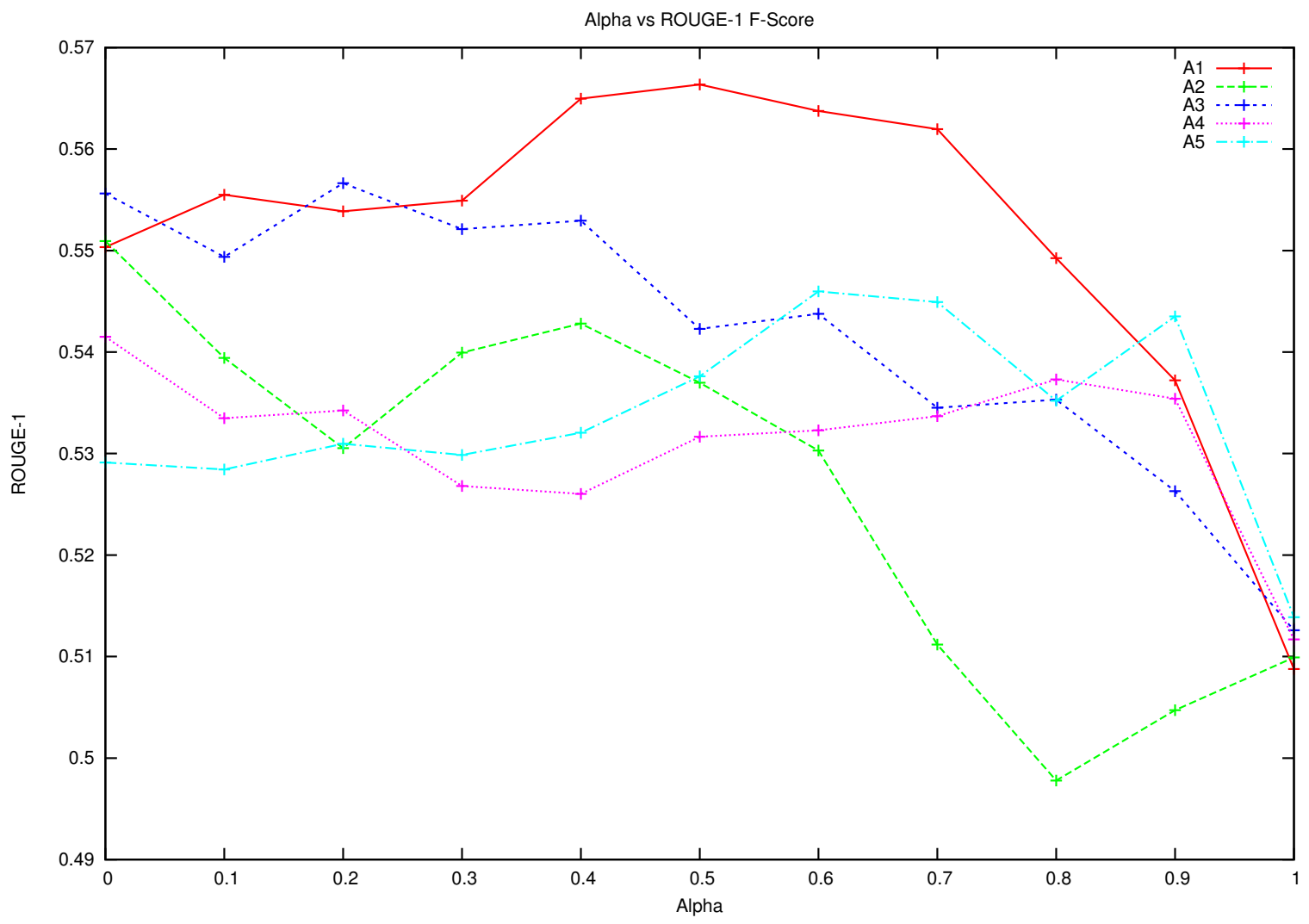

Figure 6.5: ROUGE 1 F-score *vs* $\alpha$ (r=0.25)

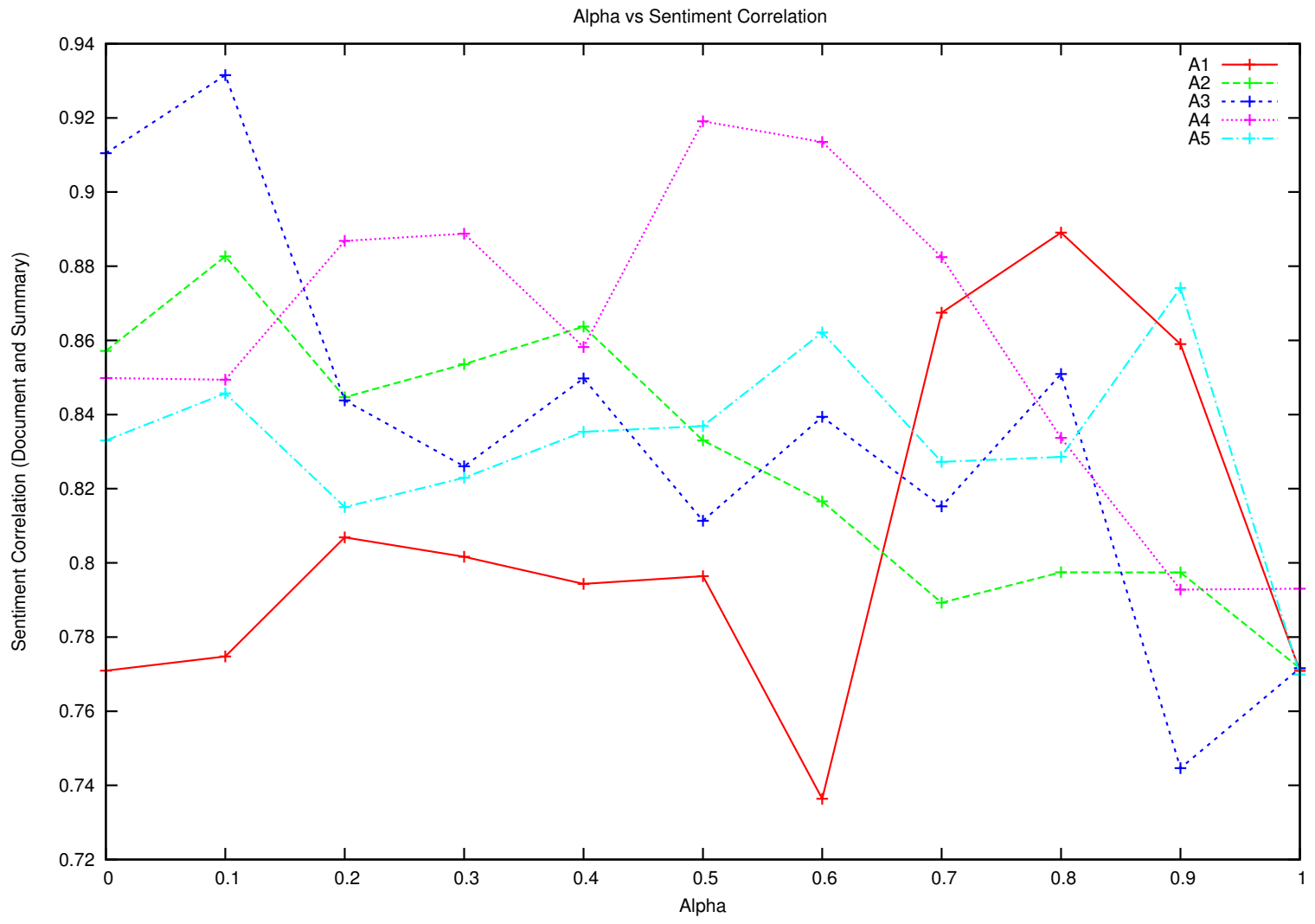

Figure 6.6: Sentiment Correlation *vs* $\alpha$ (r=0.5)

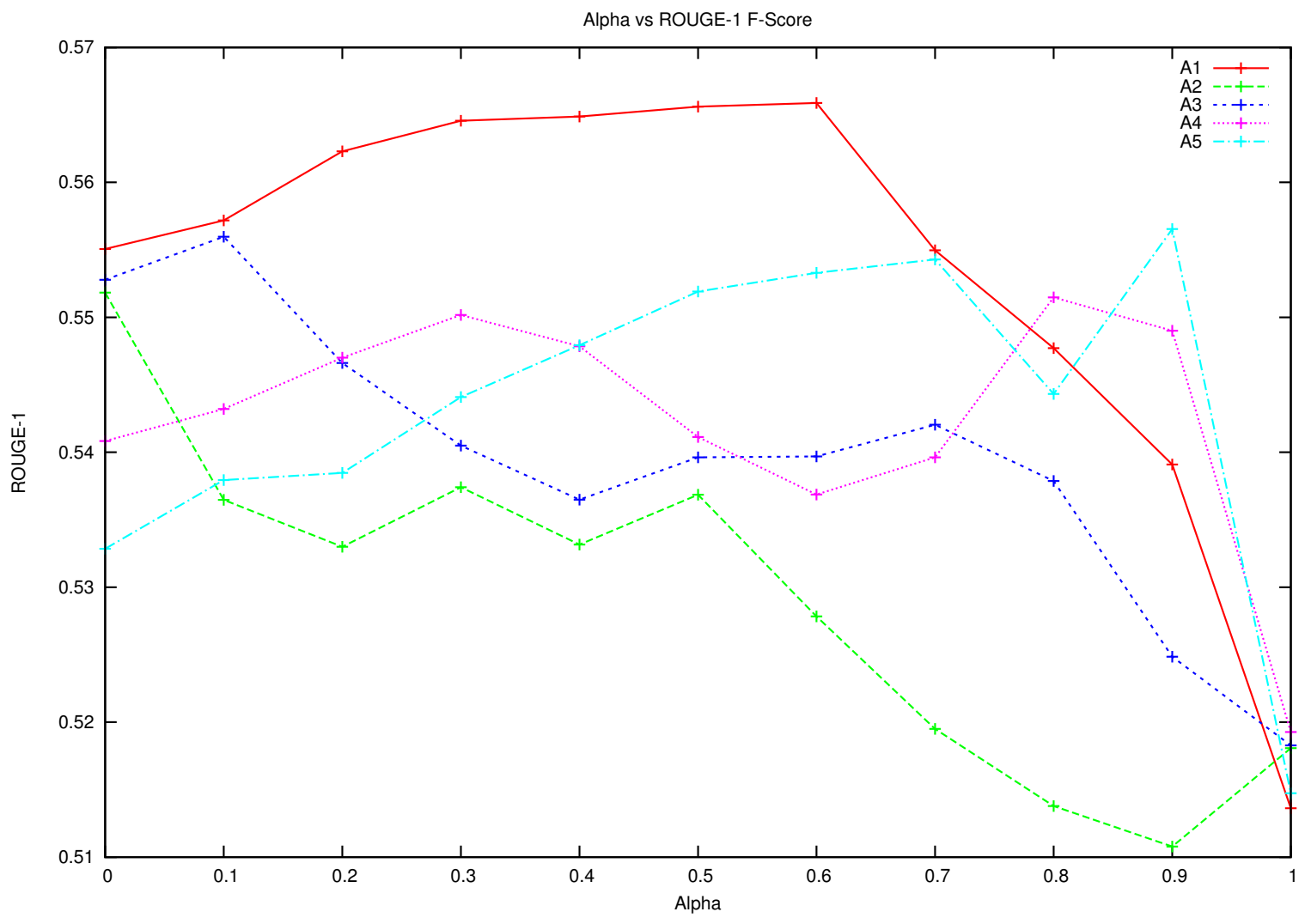

Figure 6.7: ROUGE 1 F-score *vs* $\alpha$ (r=0.5)

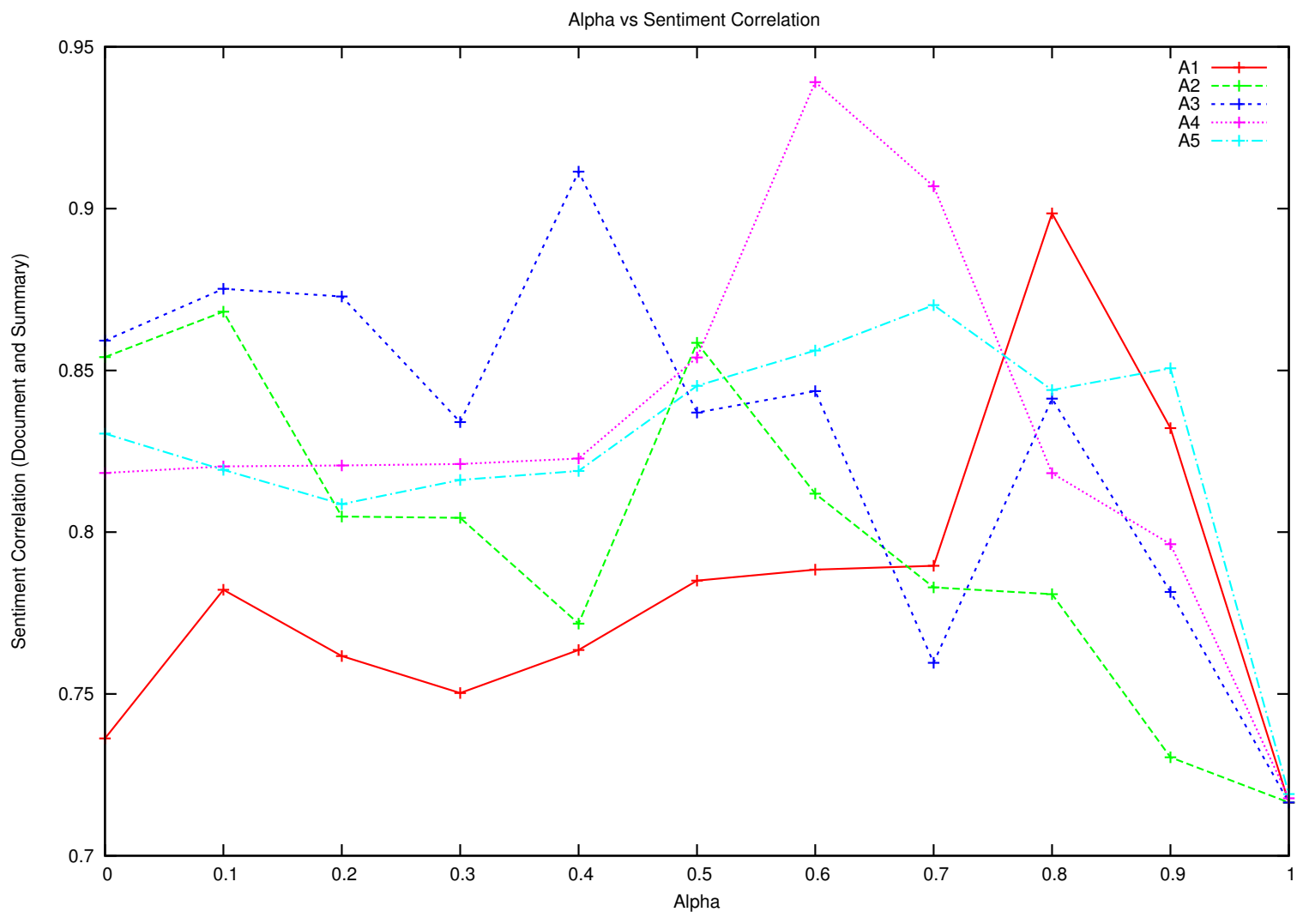

Figure 6.8: Sentiment Correlation *vs* $\alpha$ (r=0.75)

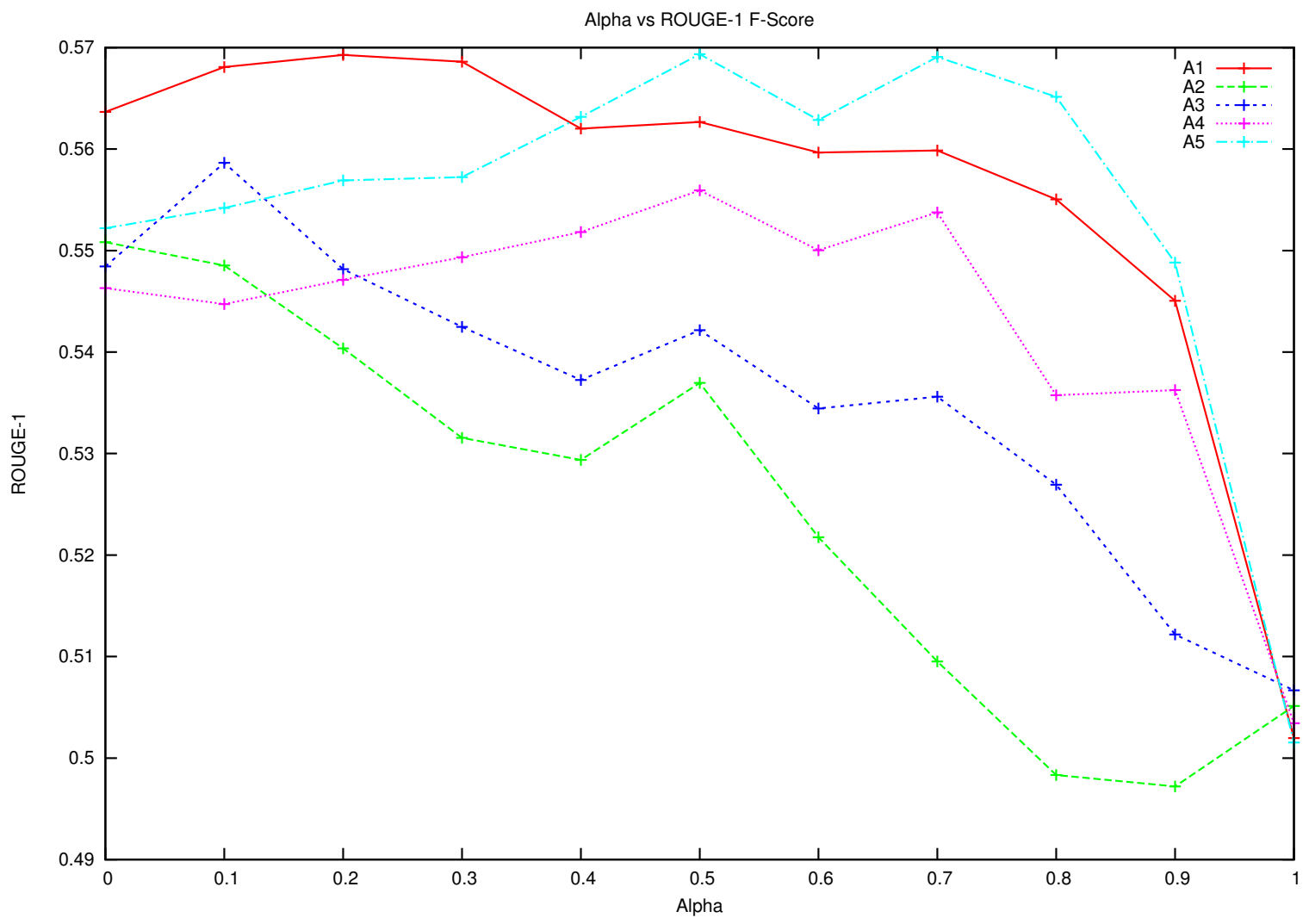

Figure 6.9: ROUGE 1 F-score *vs* $\alpha$ (r=0.75)

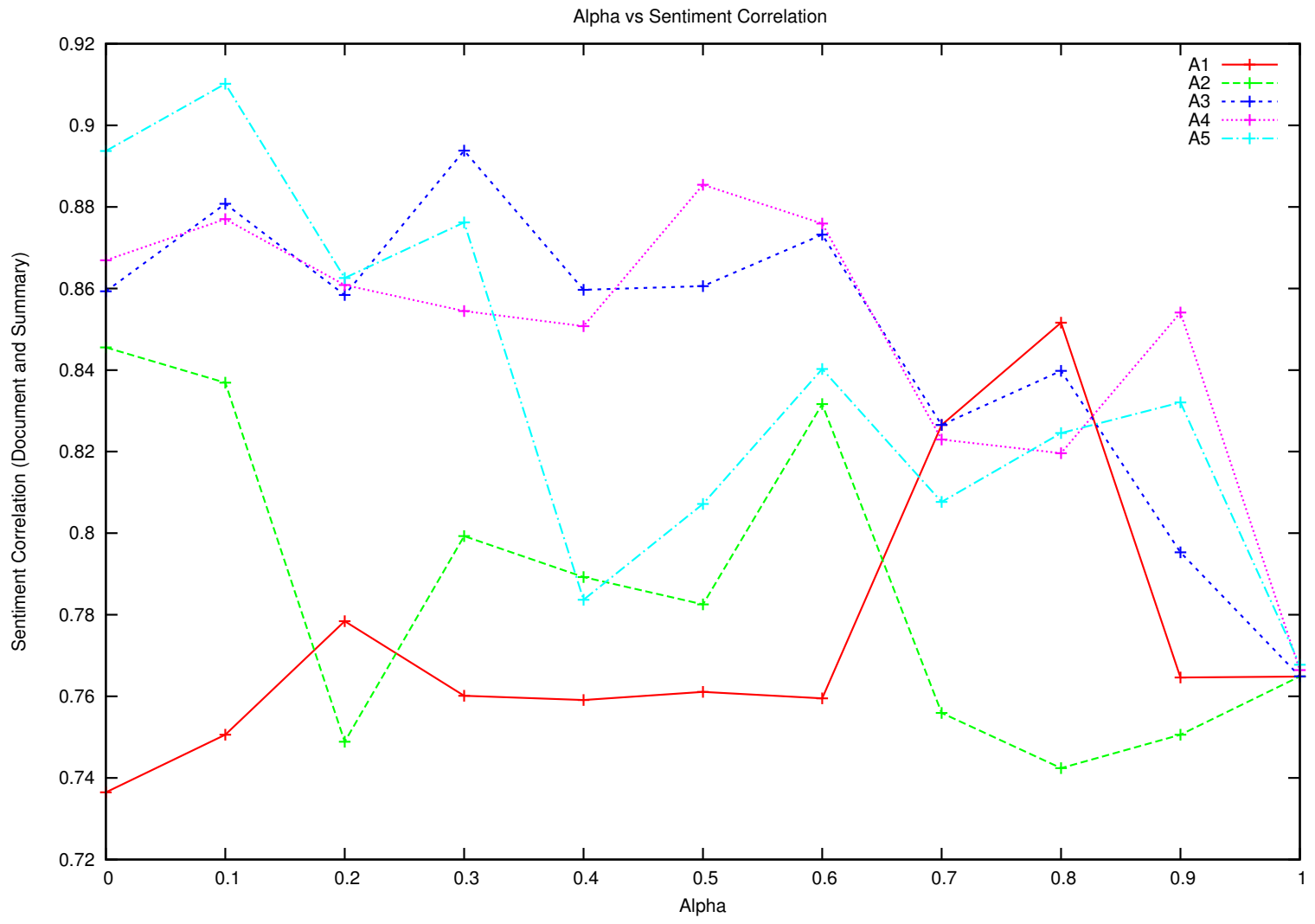

Figure 6.10: Sentiment Correlation *vs* $\alpha$ (r=1)

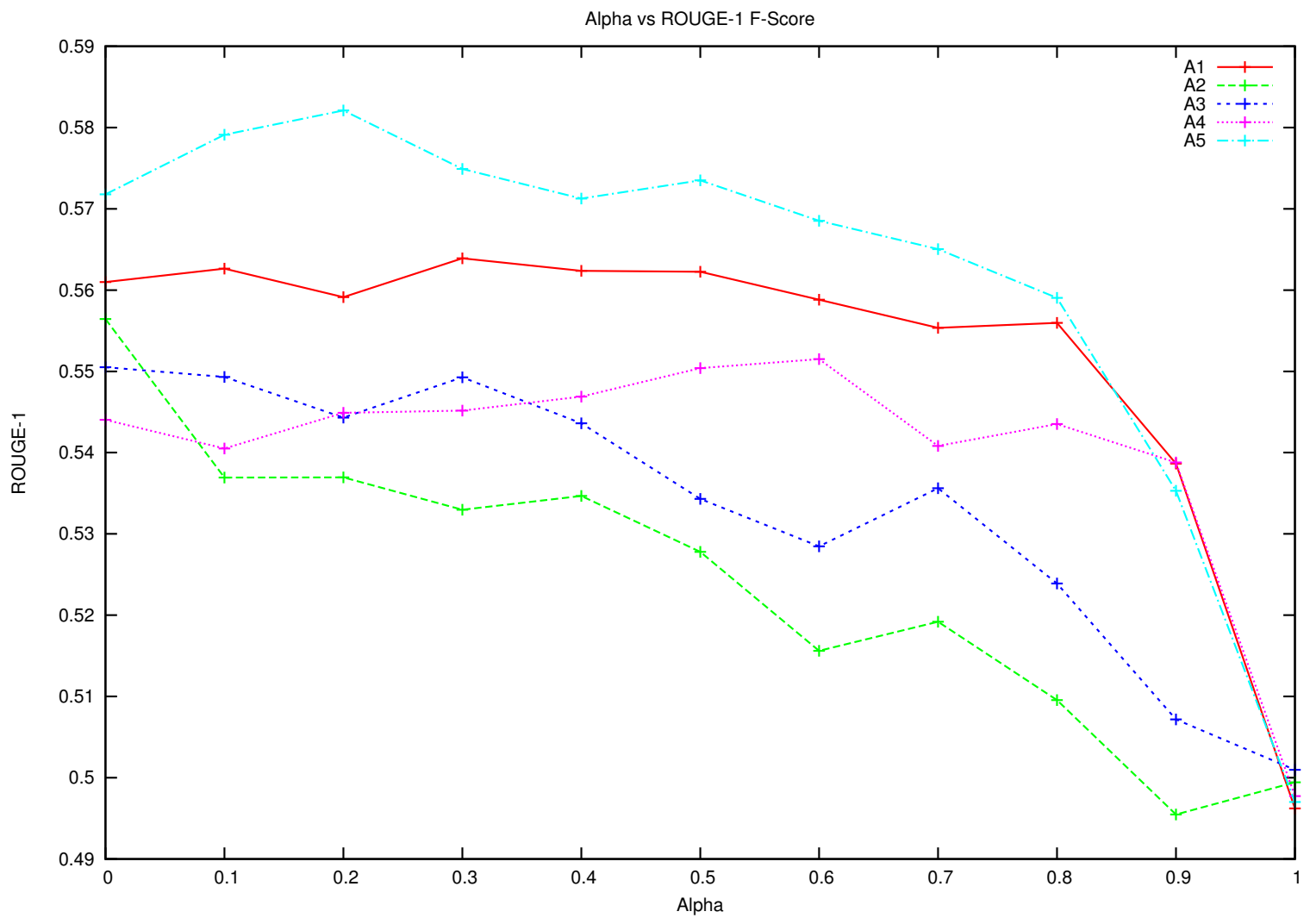

Figure 6.11: ROUGE 1 F-score *vs* $\alpha$ (r=1)

vation is calculated as the correlation between the sentiment score of the document and the corresponding summary sentiment, both calculated by the Naive Bayes sentiment classifier while the measure of coverage of information content is given by ROUGE-1 and ROUGE-2 f-scores.

Following five baselines are used for comparison:

- **Baseline-1/TOP** - Sentences selected consecutively from the start of the review within the budget.

- **Baseline-2/TOP-SUBJ** - Sentences ranked based on their subjectivity and then, selected with in the budget.

- **Baseline-3/LERMAN-SM** - [14] Sentences which have sentiment close to document sentiment are chosen as Summary. We have used same NaiveBayes classsifier [33] trained on imdb corpus to predict the sentiment of a sentence and document.

$$min_{s \subset D} \sum_{j \in s} (|senti(D) - senti(j)|) \tag{6.14}$$

- **Baseline-4/TEXTRANK** - TextRank summarizer is based on Graph based unsupervised algorithm. Graph is constructed by creating a vertex for each sentence in the document and edges between vertices based on the number of words two sentences (of vertices) have in common and then, ranking them by applying PageRank to the resulting graph. Summary is generated with sentences having more vertex score [25].

| System | ROUGE -1 | ROUGE -2 | Senti. Corr. |
|---|---|---|---|
| TOP | 0.43001 | 0.16591 | 0.86144 |
| TOP-SUBJ | 0.41807 | 0.14362 | 0.82953 |
| LERMAN-SM | 0.42608 | 0.14533 | 0.96545 |
| TEXTRANK | 0.41987 | 0.14644 | 0.88967 |
| MINCUT | 0.39368 | 0.11047 | 0.84017 |
| Our System - $A_1$ | 0.43223 | 0.15702 | 0.95306 |
| Our System - $A_2$ | 0.43594 | 0.15977 | 0.97538 |
| Our System - $A_3$ | 0.43247 | 0.15436 | 0.93155 |
| Our System - $A_4$ | **0.43602** | **0.15760** | **0.98566** |
| Our System - $A_5$ | 0.42976 | 0.15551 | 0.95415 |

Table 6.1: ROUGE F-score and sentiment correlation for optimal values of tradeoff, $\alpha$ and scaling factor, r

- **Baseline-5/MINCUT** - Mincut algorithm [31] classifies the sentences as subjective and objective sentences, by finding minimum s-t cuts in graph of sentences using maximum flow algorithm. In the graph, each sentence is a vertex and the edge between the vertex to the source or sink is taken as probability of the sentence being subjective or objective (individual scores). To ensure the graph connectivity, edges are drawn between every pair of sentence vertices, with edge weights taken proportional to the degree of proximity (association scores). After maximum flow algorithm, the cut in which source vertex lies is classified as subjective and vice-versa. We pick top subjective sentences within the budget as summary.

Table 6.1 compares the five functions with the above baselines based on optimal values of tradeoff, and scaling factor, r [18]). The table clearly shows that each of the functions outperforms the-state-of-the-art algorithms. From the table, it can be inferred that all the proposed functions not only outperform the baselines in terms of ROUGE scores for optimal parameters but also, give better correlation with the document sentiment. The main reason being that the functions with optimal values of trade-off parameter $\alpha$ strike out a balance between relevance and subjectivity. Clearly, the facility location based monotone submodular functions are the best choice as objective for opinion summarization task as they select sentences with maximum subjectivity (facilities giving maximum utility).

| System | ROUGE -1 | ROUGE -2 | Senti. Corr. |
|---|---|---|---|
| $A_1$ | 0.43223 | 0.15702 | 0.84827 |
| $A_2$ | 0.43594 | 0.15977 | 0.88601 |
| $A_3$ | 0.43247 | 0.15436 | 0.87038 |
| $A_4$ | 0.43602 | 0.15760 | 0.87818 |
| $A_5$ | 0.42976 | 0.15551 | 0.90147 |

Table 6.2: Maximum ROUGE F-score and their corresponding sentiment correlation

| System | ROUGE -1 | ROUGE -2 | Senti. Corr. |
|---|---|---|---|
| $A_1$ | 0.42572 | 0.14939 | 0.95306 |
| $A_2$ | 0.41764 | 0.14836 | 0.97538 |
| $A_3$ | 0.42415 | 0.14782 | 0.93155 |
| $A_4$ | 0.42492 | 0.14942 | 0.98566 |
| $A_5$ | 0.42572 | 0.14266 | 0.95415 |

Table 6.3: Maximum sentiment correlation and corresponding ROUGE F-Score

Looking at figure 2.3, we can observe that more weightage to relevance over subjective coverage of aspects decreases the sentiment correlation, which was expected because the summary generated misses out on subjective sentiment due to trade-off. Similarly, by looking at the figure 5.1, we also observe that more weightage to relevance over subjective coverage of aspects increases the ROUGE score as expected. The table 6.4 presents the value of sentiment correlation corresponding to maximum ROUGE score (for $\alpha \approx 1$) and table 6.3 presents the value of ROUGE score corresponding to maximum sentiment correlation (for $\alpha \approx 0$).

Both curves along with tables justifies our prior notion that relevance and subjectivity conflict in the task of opinion summarization. Furthermore, the graphs provide another method to find optimal trade-off parameter $\alpha$, by superimposing the two curves generated using extensive search, fitted as per scale such that both subjectivity as well as relevance contribute optimally in the summary.

## 6.8 Case Study

For sentences in the document, there are following cases possible:

| **Present in** | **summary with $\alpha = 1$** | **summary with $\alpha = 0$** |
|---|---|---|
| Opinion Summary | ✓ | ✓ |
| Summary | ✓ | × |
| Sentiment | × | ✓ |
| Neither | × | × |

Table 6.4: Possible presence of sentence

Consider the following movie review from Mumbai Mirror:

"*It is disconcerting to suddenly be told that ghosts don't always result in horror films. In India, they almost never do, because we end up laughing at the screen anyway. Perhaps the term 'horror comedy' generates from here. This is also the self-referential humour Bhoothnath Returns, the sequel to Bhoothnath, excels at.*

*The initial plot is propped up by cheeky innovation: Mr. Bhoothnath begs his Superior up in Bhoot World (where choice of rebirth as Aamir Khan's dog takes only one week)âĂŤwhich intentionally resembles Europe (First World= Heaven)âĂŤto give him another chance to prove his scariness. Once he is sent back down to Earth to frighten more kids, another technical glitch in the ghostly government system means that he can be seen exclusively by a kid (Bhalerao)*

*from Dharavi. This is where the relevance of the title ends. Director Nitesh Tiwari triples his ambition and makes Mr. Bhoothnath play Morgan Freeman to run for Elections, but makes sure the experience remains as educative as possible. Given the caliber of politicians that run the nation, a ghost yearning to be a political leader seems convincing and viableâĂŤmuch like a gangster hallucinating and taking advice from Mahatma Gandhi. This only goes to show how the sensibilities of an audience are molded by their environment's economic state.*

*This is a film that could have fallen apart into cliched pieces if not for its painstakingly appropriate casting. The first half belongs to the fearless Parth Bhalerao as AkhrotâĂŤone of the finest child actors in recent times. He delivers his lines with charming nonchalance, like a 10-year version of Munna from Rangeela, due to which his side-business of vacating stubborn ghosts from deserted properties seems like a legitimate profession. Sanjay Mishra as Bhoothnath's lawyer and Boman Irani as the crooked opposition leader (sequel to his Khosla Ka Ghosla rendition) almost make this idealistic world look plausible.*

*The second half, where Bhoothnath turns into the conscience of the largest democracy on the planet, reminds me of Aarakshan and Baghban. I'm fond of neither, because of how aggressively self-righteous their messages are. But this is a film where this lack of subtlety fits in, much in line with the political blitzkrieg that has filled our surroundings lately. When a universal message is to be delivered, it is important to cater to the most superficial levels of readability. The writers take a strong but preachy stand, stretching the medium to communicate their opinion about the importance of votingâĂŤone that I don't necessarily agree with. I'd still pick this film over sanctimonious party campaigns, because when Amitabh Bachchan speaks (even as a dead man), you listen. Throw in a couple of superstar cameos and voting trivia, and intentions meld into relevance.*

*Boosted by its strategically-timed release, Tiwari sculpts an unabashed tribute to Rajkumar Hirani's genre by capturing real moods within manipulated capsules of documentation. At 160 minutes, Bhoothnath Returns feels too long, but the makers will have you believe that 'Change takes time'.*

*My parents still call me up whenever Baghban plays on television; they feel well-represented by its contrivances and simplicity. If this film has half that effect, it'd have served a sincere purpose. In Bhoothnath's self-regulating words though: This is not popularity. This is curiosity.*
"

Here is 200 word summary generated by using A5 as optimization function and $\alpha$ set to one (high relevance):

*"The initial plot is propped up by cheeky innovation: Mr. Bhoothnath begs his Superior up in Bhoot World (where choice of rebirth as Aamir Khan's dog takes only one week)âĂŤwhich intentionally resembles Europe (First World= Heaven)âĂŤto give him another chance to prove his scariness. Once he is sent back down to Earth to frighten more kids, another technical glitch in the ghostly government system means that he can be seen exclusively by a kid (Bhalerao) from Dharavi. Director Nitesh Tiwari triples his ambition and makes Mr. Bhoothnath play Morgan Freeman to run for Elections, but makes sure the experience remains as educative as possible. He delivers his lines with charming nonchalance, like a 10-year version of Munna from Rangeela, due to which his side-business of vacating stubborn ghosts from deserted properties seems like a legitimate profession. I'm fond of neither, because of how aggressively self-righteous their messages are. The writers take a strong but preachy stand, stretching the medium to communicate their opinion about the importance of votingÃćâĆňâĂİone that I don't necessarily agree with. If this film has half that effect, it'd have served a sincere purpose. This is curiosity. "*

Here is 200 word summary generated by using A5 as optimization function and $\alpha$ set to zero (high subjectivity):

*"It is disconcerting to suddenly be told that ghosts don't always result in horror films. In India, they almost never do, because we end up laughing at the screen anyway. Perhaps the term 'horror comedy' generates from here. This is also the self-referential humour Bhoothnath Returns, the sequel to Bhoothnath, excels at. This is where the relevance of the title ends. This only goes to show how the sensibilities of an audience are molded by their environment's economic state. This is a film that could have fallen apart into cliched pieces if not for its painstakingly appropriate casting. I'm fond of neither, because of how aggressively self-righteous their messages are. But this is a film where this lack of subtlety fits in, much in line with the political blitzkrieg that has filled our surroundings lately. Throw in a couple of superstar cameos and voting trivia, and intentions meld into relevance. My parents still call me up whenever Baghban plays on television; they feel well-represented by its contrivances and simplicity. If this film has half that effect, it'd have served a sincere purpose. In Bhoothnath's self-regulating words though: This is not popularity. This is curiosity. "*

Consider the sentence which is present in both the summaries: "*If this film has half that effect, it'd have served a sincere purpose.*" It is not only relevant but also contain sentiment bearing words like *sincere*.

Consider the sentence which is present in only the summary with high relevance: "*It is disconcerting to suddenly be told that ghosts don't always result in horror films.* " It is a relevant sentence. Here, it also contains a sentiment bearing word - *disconcerting*.

Consider the sentence which is present in in only the summary with high subjectivity: "*He delivers his lines with charming nonchalance, like a 10-year version of Munna from Rangeela, due to which his side-business of vacating stubborn ghosts from deserted properties seems like a legitimate profession.*" The above sentence contains a lot of sentiment bearing words like *charming, stubborn etc.*

Consider the sentence which is present in neither of the summaries: "*At 160 minutes, Bhoothnath Returns feels too long, but the makers will have you believe that 'Change takes time'.* " This sentence doesnot contain any explicit sentiment bearing words nor does it contain relevant information.

## 6.9 System

OSum, Opinion Summarization Tool is designed for general use of the movie review summary generation tool. It is hosted on CFILT server[1]. Given a movie review along with parameters to the function, it will generate the summary. The figure 6.12 and 6.13 show its working. Figure 6.12 shows the input page while Figure 6.13 shows the output page.

### 6.9.1 User Interface Layout

The interface of the Opinion Summarization Tool consists of the following I/O features. The input to the interface consists of:

- Drop box to select one out of five functions
- Text box with linked slider to select the value of trade-off, $\alpha$

[1]http://10.144.22.120:8040/os/

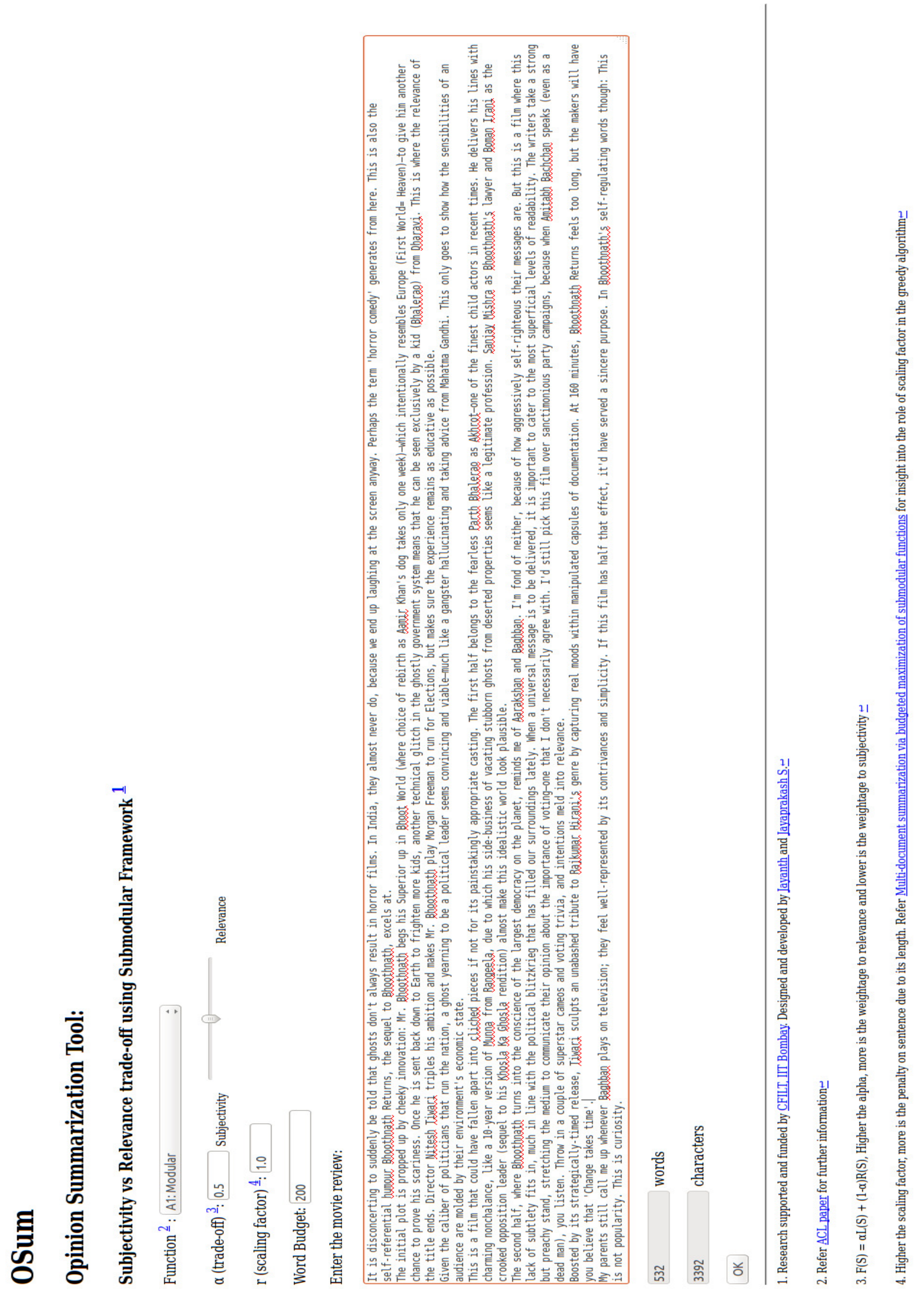

Figure 6.12: Opinion Summarization System - Input Page

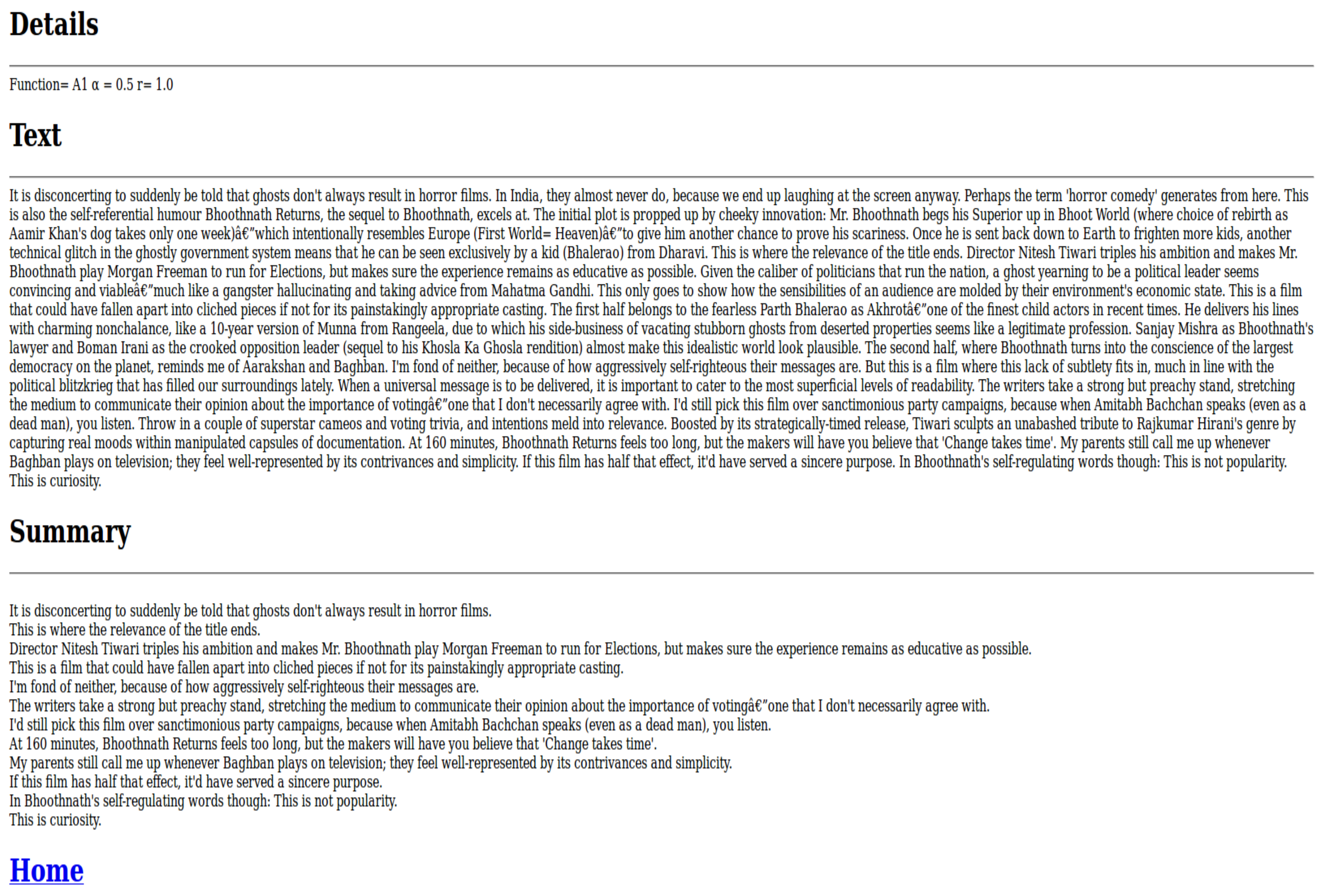

# Details

Function= A1 α = 0.5 r= 1.0

## Text

It is disconcerting to suddenly be told that ghosts don't always result in horror films. In India, they almost never do, because we end up laughing at the screen anyway. Perhaps the term 'horror comedy' generates from here. This is also the self-referential humour Bhoothnath Returns, the sequel to Bhoothnath, excels at. The initial plot is propped up by cheeky innovation: Mr. Bhoothnath begs his Superior up in Bhoot World (where choice of rebirth as Aamir Khan's dog takes only one week)â€"which intentionally resembles Europe (First World= Heaven)â€"to give him another chance to prove his scariness. Once he is sent back down to Earth to frighten more kids, another technical glitch in the ghostly government system means that he can be seen exclusively by a kid (Bhalerao) from Dharavi. This is where the relevance of the title ends. Director Nitesh Tiwari triples his ambition and makes Mr. Bhoothnath play Morgan Freeman to run for Elections, but makes sure the experience remains as educative as possible. Given the caliber of politicians that run the nation, a ghost yearning to be a political leader seems convincing and viableâ€"much like a gangster hallucinating and taking advice from Mahatma Gandhi. This only goes to show how the sensibilities of an audience are molded by their environment's economic state. This is a film that could have fallen apart into cliched pieces if not for its painstakingly appropriate casting. The first half belongs to the fearless Parth Bhalerao as Akhrotâ€"one of the finest child actors in recent times. He delivers his lines with charming nonchalance, like a 10-year version of Munna from Rangeela, due to which his side-business of vacating stubborn ghosts from deserted properties seems like a legitimate profession. Sanjay Mishra as Bhoothnath's lawyer and Boman Irani as the crooked opposition leader (sequel to his Khosla Ka Ghosla rendition) almost make this idealistic world look plausible. The second half, where Bhoothnath turns into the conscience of the largest democracy on the planet, reminds me of Aarakshan and Baghban. I'm fond of neither, because of how aggressively self-righteous their messages are. But this is a film where this lack of subtlety fits in, much in line with the political blitzkrieg that has filled our surroundings lately. When a universal message is to be delivered, it is important to cater to the most superficial levels of readability. The writers take a strong but preachy stand, stretching the medium to communicate their opinion about the importance of votingâ€"one that I don't necessarily agree with. I'd still pick this film over sanctimonious party campaigns, because when Amitabh Bachchan speaks (even as a dead man), you listen. Throw in a couple of superstar cameos and voting trivia, and intentions meld into relevance. Boosted by its strategically-timed release, Tiwari sculpts an unabashed tribute to Rajkumar Hirani's genre by capturing real moods within manipulated capsules of documentation. At 160 minutes, Bhoothnath Returns feels too long, but the makers will have you believe that 'Change takes time'. My parents still call me up whenever Baghban plays on television; they feel well-represented by its contrivances and simplicity. If this film has half that effect, it'd have served a sincere purpose. In Bhoothnath's self-regulating words though: This is not popularity. This is curiosity.

## Summary

It is disconcerting to suddenly be told that ghosts don't always result in horror films.
This is where the relevance of the title ends.
Director Nitesh Tiwari triples his ambition and makes Mr. Bhoothnath play Morgan Freeman to run for Elections, but makes sure the experience remains as educative as possible.
This is a film that could have fallen apart into cliched pieces if not for its painstakingly appropriate casting.
I'm fond of neither, because of how aggressively self-righteous their messages are.
The writers take a strong but preachy stand, stretching the medium to communicate their opinion about the importance of votingâ€"one that I don't necessarily agree with.
I'd still pick this film over sanctimonious party campaigns, because when Amitabh Bachchan speaks (even as a dead man), you listen.
At 160 minutes, Bhoothnath Returns feels too long, but the makers will have you believe that 'Change takes time'.
My parents still call me up whenever Baghban plays on television; they feel well-represented by its contrivances and simplicity.
If this film has half that effect, it'd have served a sincere purpose.
In Bhoothnath's self-regulating words though: This is not popularity.
This is curiosity.

## Home

Figure 6.13: Opinion Summarization System - Output Page

- Text box to select the value of scaling parameter, $r$
- Text box to select the value of word budget, which is less than or equal to the word count
- Text area to input the movie review text, which initially contains a default review text
- Read only box giving count of words in the text area
- Read only box giving count of characters in the text area
- OK button to submit the text for summarization

The output to the interface consists of:

- Details of the parameters
- Input text
- Generated summary of the input text

## 6.10 Summary

In this chapter, we have justified the submodular property of opinion summary through examples and good performance of the system over the baselines. Another interesting message inferred from the experiments is that subjectivity has to be given more weightage in opinion summary over relevance as ROGUE Score eventually decreases after increasing weightage to relevance. This can be attributed to the general trend in *NLP*, where higher layer *i.e.* semantics, pragmatics and discourse eventually dominates over lower layers in cases of conflict. Summary is based on lexical and syntactic extraction while subjectivity is based on semantics and pragmatics. And thus, here also the higher layers based subjectivity extraction dominates over relevance.

# Chapter 7

# Conclusion and Future Work

As we reach to the end of this report, we conclude our work and propose some future work related to addressing the problems faced or improving the system.

## 7.1 Conclusion

Doucument summarization problem is one of the well known problems in NLP. Summary of document could be a set of keywords, set of sentences which covers differents features of original documents like relevancy, sentiment, aspect coverage, etc.

For Keyword extraction and Image retrieval, we have used the probablisitic models for caption generation. Though caption generation did not yield good result as individual system, When we use this in rank aggregation framework which combines results of various systems. And also it is observed that both manual evaluation and automatic evaluation results improved over existing systems. (P@1 $\approx$ 2.5%, P@5 $\approx$ 4%).

In chapter 6, we have justified the submodular property of opinion summary through examples and good performance of the system over the baselines. And also through experiments we have shown that there exist a trade-off between choosing relevance sentence and sentiment bearing sentences.

## 7.2 Future work

Future work of the project has been divided in two subsections depending on their complexity and importance to the project.

### 7.2.1 Short Term

#### Clustering Terms

It is observed that even the keywords extracted from are articles actually important and informative, however same combinations of keywords may not exist at image collection side. Clustering may be helpful, if we could cluster the terms and pick the terms from a cluster.

### 7.2.2 Long Term

#### Semantic Understanding of Document

Long term goal is moving beyond surface level keyword extraction. Sometimes set of keyword extracted alone doesnot depict actual meaning. Trying to comeup with a way of representing a text document as graph and relationships among concepts, may help choosing better words for formulating query.